\documentclass[aps,amsmath,amssymb,nofootinbib,preprintnumbers,showpacs]{revtex4}
\usepackage{epsfig}
\usepackage{bbm}
\usepackage{mathrsfs,graphicx,bm,amsmath}

\newcommand{\beq}{\begin{equation}}
\newcommand{\eeq}{\end{equation}}

\newcommand{\STr}{\mathrm{STr}}

\newcommand{\lit}{^6\mathrm{Li}}
\newcommand{\kal}{^{40}\mathrm{K}}

\def\di{\displaystyle}

\def\bg{\begin{eqnarray}\begin{array}{rcl}\displaystyle}
\def\eg{\end{array} &\di    &\di   \end{eqnarray}}
\def\bm#1{\begin{eqnarray}\begin{array}{#1}\di}
\def\bmo#1{\begin{eqnarray*}\begin{array}{#1}\di}
\def\bml#1#2{\begin{eqnarray}\begin{array}{#1}\label{#2}\di}
\def\bgo{\begin{eqnarray*}\begin{array}{rcl}\displaystyle}
\def\ego{\end{array} &\di    &\di \nonumber  \end{eqnarray*}}

\def\btensor#1#2{\renew\left#1\begin{array}{#2}\di}
\def\brtensor#1#2#3{\ren#3\left#1\begin{array}{#2}}
\def\botensor#1#2{\renew\left#1\begin{array}{#2}}
\def\etensor#1{\end{array}\right#1}

\def\STr{{\rm STr}}

\def\s0#1#2{\mbox{\small{$ \frac{#1}{#2} $}}}
\def\0#1#2{\frac{#1}{#2}}

\begin{document}


\title{Efimov effect from functional renormalization}

\author{ S. Moroz\footnote{S.Moroz@thphys.uni-heidelberg.de}, S. Floerchinger\footnote{S.Floerchinger@thphys.uni-heidelberg.de}, R. Schmidt\footnote{R.Schmidt@thphys.uni-heidelberg.de}, and C. Wetterich\footnote{C.Wetterich@thphys.uni-heidelberg.de}}
\address{
Institut f{\"u}r Theoretische Physik,
Philosophenweg 16, D-69120 Heidelberg, Germany}

\today

\begin{abstract}
We apply a field-theoretic functional renormalization group technique to the few-body (vacuum) physics of non-relativistic atoms near a Feshbach resonance. Three systems are considered: one-component bosons with a $U(1)$ symmetry, two-component fermions with a $U(1)\times SU(2)$ symmetry and three-component fermions with a $U(1) \times SU(3)$ symmetry. We focus on the scale invariant unitarity limit of an infinite scattering length. The exact solution for the two-body sector is consistent with the unitary fixed point behavior of the considered systems. Nevertheless, the numerical three-body solution in the s-wave sector develops a limit cycle scaling in case of $U(1)$ bosons and $SU(3)$ fermions. The Efimov parameter for the one-component bosons and the three-component fermions is found to be $s\approx 1.006$, consistent with the result of Efimov.
\end{abstract}


\maketitle


\section{Introduction}
\label{Intro}
The physics of ultracold atoms is a broad area of research which develops rapidly both experimentally and theoretically (for reviews see \cite{giorgini:2007,bloch:2007}). To a large extend this is due to the excellent tunability and control of the studied systems. In particular the interaction strength of atoms near a Feshbach resonance can be changed in broad ranges by tuning the magnetic field, which makes these systems an ideal playground for testing the predictions of theoretical models at strong coupling. Both, few-body and many-body quantum, and thermodynamic effects have been extensively studied with ultracold gases.

Near a broad Feshbach resonance the atom-atom s-wave scattering length $a$ is large in comparison to the effective range $r_{\mathrm{eff}}$ of the microscopic interaction. The low-energy vacuum physics (for vanishing temperature and density) becomes universal: some physical observables become insensitive to the detailed form of the microscopic interaction and depend only on the scattering length $a$ \cite{Braaten04}. For example, for $a>0$ the theory admits a stable shallow diatom. For this atom-atom bound state the universal binding energy is determined simply by dimensional analysis. In the unitarity limit all energy scales drop out of the problem and the theory is scale invariant in the two-body sector. It is a well-established result, derived first by Efimov \cite{Efimov70}, that in the three-body sector of the resonantly interacting particles a spectrum of shallow three-body bound states develops. At unitarity, the spectrum is geometric which is a signature of the limit cycle behavior of the renormalization group flow. Even in the case of a scale symmetry in the two-body sector, the running of the renormalized three-body couplings indicates a violation of the dilatation symmetry and may be associated with a quantum anomaly \cite{Camblong}. 

The low-energy few-body scattering of atoms has been investigated using various computational non-perturbative techniques ranging from effective field theory \cite{Bedaque1, Bedaque2, Kagan05,Gurarie06} to quantum mechanics \cite{Efimov70,Petrov03}. The perturbative $\epsilon$ expansion around critical $d=4$ and $d=2$ dimensions has also been applied to this problem \cite{Rupak06, Nishida:2006}. A field theoretical functional renormalization group approach has been used to investigate the two-body and three-body sectors of two-component fermions recently \cite{DKS, Birse}. As a convenient truncation in vacuum, the authors use a vertex expansion and reproduce the Skorniakov and Ter-Martirosian integral equation \cite{Skorniakov56}. In this way, the universal ratio of the atom-diatom to the atom-atom scattering length is computed.\\

In this work we follow \cite{DKS} and consider the few-body physics of non-relativistic atoms near a Feshbach resonance which may be described by a simple two-channel model of particles with short-range interactions. We study three different systems: bosons with a $U(1)$ symmetry (System I), fermions with a $U(1)\times SU(2)$ symmetry (System II) and fermions with a $U(1) \times SU(3)$ symmetry (System III). Both, Systems I and II have been well-studied during the last decade. The model of $SU(3)$ fermions might be of relevance for three-component mixtures of $\lit$ atoms near the broad Feshbach resonances. The many-body properties of this model have been studied in \cite{Rapp:2006, He:2006, Brauner:2007, Paananen, Cherng}. Recently, the three-component fermion system has been studied with functional renormalization group methods using an approximation including a trion field \cite{FSMW}. The present work, which is based on a vertex expansion, complements and extends the results of \cite{FSMW}. It underlines the basic finding of the presence of Efimov states for $SU(3)$ fermions and estimates the universal Efimov parameter $s$ with a higher precision.

The structure of the paper is as follows: In Section \ref{Method} we present a field-theoretic renormalization group (RG) method, which we use to solve the few-body problem, and introduce the three models we are going to investigate in this work. In Section \ref{VacPro} we investigate the effective action in the vacuum state, i.e. for vanishing temperature $T=0$ and density $n=0$. The following Section \ref{twobody} is devoted to the exact solution of the two-body sector for positive scattering lengths $a>0$ (diatom phase). In Section \ref{Threeflow} we turn to the analysis of the three-body sector and derive the RG flow equation for the atom-diatom vertex at unitarity. This RG equation is solved analytically employing a simple pointlike approximation in Section \ref{triatom}. In Sections \ref{one-two} and \ref{thg} we reproduce the Skorniakov and Ter-Martirosian integral equation and present a numerical solution of the three-body RG flow equation. We draw our conclusions in Section \ref{Conclusion}.  

\section{Method and definition of models} 
\label{Method}
In this work we calculate a scale-dependent effective action functional $\Gamma_{k}$ \cite{Wetterich:1992yh} (for reviews see \cite{Berges:2000ew,Aoki:2000wm}), often called average action, flowing action, or running action. This renormalization group method is formulated in Euclidean spacetime using the Matsubara formalism. The flowing action $\Gamma_{k}$ includes all fluctuations with momenta $q\gtrsim k$. In the infrared limit $k\to 0$ the full effective action $\Gamma=\Gamma_{k\to 0}$ is obtained. This dependence on the scale $k$ is introduced by adding a regulator $R_{k}$ to the inverse propagator $\Gamma^{(2)}_{k}$and the flowing action $\Gamma_k$ obeys the exact functional flow equation \cite{Wetterich:1992yh}:
\begin{eqnarray}\label{M1}
  \partial_k \Gamma_k &=& \frac{1}{2} \STr \,
  \partial_k R_k \, (\Gamma^{(2)}_k + R_k)^{-1} = \frac{1}{2} \STr\,\tilde \partial_k \,\ln
  (\Gamma^{(2)}_k + R_k).
\end{eqnarray}
This functional differential equation for $\Gamma_{k}$ must be supplemented with the initial condition $\Gamma_{k\to\Lambda}=S$, where the ``classical action'' $S$ describes the physics at the microscopic UV scale, $k=\Lambda$. In Eq. (\ref{M1}) $\STr$ denotes a supertrace which sums over momenta, Matsubara frequencies, internal indices, and fields (taking fermions with a minus sign). The second functional derivative $\Gamma^{(2)}_{k}$ denotes the full inverse field propagator, which is modified by the presence of the IR regulator $R_k$. As a consequence, the fluctuations with $q^2<k^2$ are suppressed and the effective action depends on the scale $k$. The choice of the momentum dependent regulator function $R_{k}(q)$ introduces a scheme dependence which has to disappear for the exact solution for $k\to 0$. In the second form of the flow equation (\ref{M1}) $\tilde\partial_k$ denotes a scale derivative, which acts only on the IR regulator $R_{k}$. This form is very useful because it can be formulated in terms of one-loop Feynman diagrams. The effective action $\Gamma_{k=0}$ is the generating functional of the 1PI vertices, which can be easily connected to the different scattering amplitudes in the case of vanishing density ($n=0$) and vanishing temperature ($T=0$). It is also convenient to introduce the RG ``time'' $t\equiv \ln(k/\Lambda)$, which flows in the interval $t\in(-\infty,0)$.  In the following we will use both $t$ and $k$.

In most cases of interest the functional differential equation (\ref{M1}) can be solved only approximately. Usually some type of expansion of $\Gamma_{k}$ is performed, which is then truncated at finite order leading to a finite system of ordinary differential equations. The expansions do not necessarily involve a small parameter (like an interaction coupling constant) and they are, in general, of non-perturbative nature. As has already been advocated in Sect. \ref{Intro}, we perform a systematic vertex expansion of $\Gamma_{k}$ taking the full momentum dependence of the relevant vertex in the three-body sector into account. The vertex expansion is an expansion in powers of fields; hence generally:
\beq \label{M2}
\Gamma_{k}=\sum_{n=0}^{\infty}\Gamma_{k}(n)=\Gamma_{k}(2)+\Gamma_{k}(3)+\Gamma_{k}(4)+...,
\eeq
where the index in brackets denotes the number of fields $n$ in the  monomial term $\Gamma_{k}(n)$. In the second equation $\Gamma_{k}(0)$ and $\Gamma_{k}(1)$ are missing because we are not interested in the free energy of the vacuum and the term linear in the fields is absent by construction.\\
 
In this paper we are interested in the  non-relativistic physics of atoms interacting via a Feshbach resonance, which can be described by a simple two-channel model. In particular, we consider and compare three different systems:
\begin{itemize}
\item System I: Single bosonic field near a Feshbach resonance

Our truncation of the scale-dependent flowing action, written in the Fourier space, is:
\begin{eqnarray} \label{M3}
\Gamma_k(2)&=&\int\limits_{Q}\psi^{*}(Q)(i\omega_{\mathbf{q}}+\mathbf{q}^2-\mu_{\psi})\psi(Q)+\int\limits_{Q}\varphi^{*}(Q)P_{\varphi}(Q)\varphi(Q) \nonumber \\
\Gamma_k(3)&=&\frac{h}{2}\int\limits_{Q_1,Q_2,Q_3}\Big[\varphi^{*}(Q_1)\psi(Q_2)\psi(Q_3)+\varphi(Q_1)\psi^{*}(Q_2)\psi^{*}(Q_3) \Big] \delta(Q_1-Q_2-Q_3) \nonumber \\
\Gamma_k(4)&=&-\int\limits_{Q_1,...Q_4}  \lambda_{3}(Q_{1},Q_{2},Q_{3})\varphi(Q_{1})\psi(Q_{2})\varphi^{*}(Q_{3})\psi^{*}(Q_{4})\delta(Q_{1}+Q_{2}-Q_{3}-Q_{4}),
\end{eqnarray} 
where $Q=(\omega, \mathbf{q})$ and $\int\limits_{Q}=\int_{-\infty}^{\infty} \frac{d\omega}{2\pi} \int_{-\infty}^{\infty} \frac{d^{3}\mathbf{q}}{(2\pi)^3}$. The field $\psi$ represents an elementary complex bosonic atom, while $\varphi(Q)$ is a complex bosonic composite diatom which mediates the Feshbach interaction. At the initial UV scale we take $\lambda_{3}=0$. The action for $\varphi$ becomes Gaussian, and one may integrate out $\varphi$ using its field equation $\varphi\sim \psi\psi$. As will be demonstrated in Sect. \ref{twobody}, the Yukawa coupling $h$ is simply related to the width of the Feshbach resonance. For $k\to 0$ the coupling $\lambda_3(Q_1,Q_2,Q_3)$ becomes the 1PI vertex which can be connected to the atom-diatom  scattering amplitude. The system has an obvious $U(1)$ symmetry which reflects the conserved number of atoms\footnote{In general, the $U(1)$ symmetry can be spontaneously broken due to many-body effects and our truncation (\ref{M3}) would be insufficient. In this work, however, we are interested only in the few-body physics (for more details see Sect. \ref{VacPro}).
}.

\item System II: Fermionic doublet near a Feshbach resonance

\begin{eqnarray} \label{M4}
\Gamma_k(2)&=&\sum_{i=1}^{2}\int\limits_{Q}\psi^{*}_{i}(Q)(i\omega_{\mathbf{q}}+\mathbf{q}^2-\mu_{\psi})\psi_{i}(Q)+\int\limits_{Q}\varphi^{*}(Q)P_{\varphi}(Q)\varphi(Q) \nonumber \\
\Gamma_k(3)&=&-h\int\limits_{Q_1,Q_2,Q_3}\Big[\varphi^{*}(Q_1)\psi_{1}(Q_2)\psi_{2}(Q_3)-\varphi(Q_1)\psi^{*}_{1}(Q_2)\psi^{*}_{2}(Q_3) \Big] \delta(Q_1-Q_2-Q_3) \nonumber \\
\Gamma_k(4)&=&\int\limits_{Q_1,...Q_4}  \lambda_{3}(Q_{1},Q_{2},Q_{3})\sum_{i=1}^{2}\varphi(Q_{1})\psi_{i}(Q_{2})\varphi^{*}(Q_{3})\psi^{*}_{i}(Q_{4}) \delta(Q_{1}+Q_{2}-Q_{3}-Q_{4}).
\end{eqnarray} 
Here, the two species of elementary fermionic atoms $\psi_{1},\psi_{2}$ are described by Grassmann-valued fields, and $\varphi$ is a composite bosonic diatom. At the UV scale one has $\varphi\sim \psi_{1}\psi_{2}$. This fermionic system has an $SU(2)\times U(1)$ internal symmetry with $(\psi_{1},\psi_{2})$ transforming as a doublet and $\varphi$ as a singlet of the $SU(2)$ flavor subgroup. Two-species fermion systems near Feshbach resonances were realized experimentally with $\lit$ and $\kal$ atoms \cite{Ultracold}.  

\item System III: Fermionic triplet near a Feshbach resonance
\begin{eqnarray} \label{M5}
\Gamma_k(2)&=&\int\limits_{Q}\sum_{i=1}^{3}\psi^{*}_{i}(Q)(i\omega_{\mathbf{q}}+\mathbf{q}^2-\mu_{\psi})\psi_{i}(Q)+\int\limits_{Q}\sum_{i=1}^{3}\varphi^{*}_{i}(Q)P_{\varphi}(Q)\varphi_{i}(Q) \nonumber \\
\Gamma_k(3)&=&\frac{h}{2}\int\limits_{Q_1,Q_2,Q_3}\sum_{i,j,k=1}^{3}\epsilon_{ijk}\Big[\varphi^{*}_{i}(Q_1)\psi_{j}(Q_2)\psi_{k}(Q_3)-\varphi_{i}(Q_1)\psi^{*}_{j}(Q_2)\psi^{*}_{k}(Q_3) \Big] \delta(Q_1-Q_2-Q_3) \nonumber \\
\Gamma_k(4)&=&\int\limits_{Q_1,...Q_4}  \Big[ \lambda_{3a}(Q_{1},Q_{2},Q_{3})\sum_{i=1}^{3}\varphi_{i}(Q_{1})\psi_{i}(Q_{2})\sum_{j=1}^{3}\varphi^{*}_{j}(Q_{3})\psi^{*}_{j}(Q_{4})+ \nonumber \\
&+& \lambda_{3b}(Q_{1},Q_{2},Q_{3})\sum_{i=1}^{3}\psi_{i}(Q_{2})\psi_{i}^{*}(Q_{4})\sum_{j=1}^{3}\varphi_{j}(Q_{1})\varphi^{*}_{j}(Q_{3}) \Big ] \delta(Q_{1}+Q_{2}-Q_{3}-Q_{4}).
\end{eqnarray} 
The three species of the elementary Grassmann-valued fermion field can be assembled into a vector $\psi=(\psi_{1},\psi_{2},\psi_{3})$. Similarly the three composite Feshbach bosonic diatoms form the vector $\varphi=(\varphi_{1},\varphi_{2},\varphi_{3})\sim(\psi_{2}\psi_{3},\psi_{3}\psi_{1},\psi_{1}\psi_{2})$. The action has an $SU(3)\times U(1)$ symmetry with $\psi$ transforming as $\mathbf{3}$, and $\varphi$ as $\mathbf{\bar{3}}$ for the $SU(3)$ flavor subgroup. Two different couplings $\lambda_{3a}$ and $\lambda_{3b}$ are allowed by the $SU(3)$ symmetry. This model might be of relevance for three-component mixtures of $\lit$ atoms. There are three distinct broad Feshbach resonances for three scattering channels near $B\approx 800\,\textrm{G}$ for $\lit$ atoms. As a first approximation we assume that the resonances for all channels are degenerate, which leads to the $SU(3)$ flavor symmetry and to the model (\ref{M5}). A stable three-component mixture of $\lit$ atoms has been recently created \cite{Jochim, OHara}. The theoretical investigation of the 3-body losses in \cite{Jochim, OHara} has been recently published \cite{Braaten, Naidon,Schmidt}. 
\end{itemize}

To unify our language for the different models considered in this paper, we refer to the elementary particles $\psi$ as atoms (and denote corresponding quantities with the subscript $\psi$), while the composite $\varphi$ is called diatom. All considered systems have Galilean spacetime symmetry, which consequences we discuss in Appendix \ref{NRCFT}. Our units are $\hbar=k_{B}=1$. Moreover we choose the energy units such that $2M_{\psi}=1$, where $M_{\psi}$ is the mass of the atom.

We should stress that $\Gamma(2)$ and $\Gamma(3)$ do not have the most general form. The most general form of the vertex expansion includes an arbitrary inverse atom propagator $P_{\psi}(Q)$ and a momentum-dependent Yukawa coupling $h(Q_1,Q_2,Q_3)$. However, due to special properties of the vacuum state (see Sect. \ref{VacPro}), $P_{\psi}(Q)$ and $h(Q_1,Q_2,Q_3)$ are not renormalized and keep their microscopic values $P_{\psi}(Q)=(i\omega_{\mathbf{q}}+\mathbf{q}^2-\mu_{\psi})$ and $h(Q_1,Q_2,Q_3)=h$ during the RG flow.
At this point it is also important to note that our vertex expansion is complete to the third order in the fields. Possible terms with four fields, which are invariant with respect to the symmetries of our models, can be found in Appendix \ref{Gamma4}. In this Appendix we also present arguments, because of which we do not include these terms in our truncation. To summarize, the properties of the two and three-body sectors, which are of the main interest in this work, can be calculated using the truncations $(\ref{M3}),(\ref{M4}),(\ref{M5})$.



\section{Vacuum Limit}
\label{VacPro}
The advantage of the method used in this paper is that it is a field-theoretical setting which permits computations for the general case of non-zero temperature ($T\ne 0$) and density ($n \ne 0$). In this work we are interested only in the scattering and the bound states of few particles in \emph{vacuum}. The projection of the effective action $\Gamma_{k=0}$ onto the vacuum state must be performed carefully and was developed in \cite{Diehl:2005ae, DKS}. Here we shortly summarize the procedure:

The vacuum projection of $\Gamma_{k=0}$ is performed as follows:
\beq \label{vac1}
\Gamma_{vac}=\lim_{k_{F}\to 0, T\to 0} \Gamma_{k=0}\Big|_{T>T_{c}(k_{F})},
\eeq
where $k_{F}=(3\pi^2 n)^{1/3}$ is a formal Fermi wave vector (defined for both bosons and fermions) and $n$ is the atom density of the system. Thus we start with the effective action at finite density and temperature. The system is then made dilute by taking limit $k_{F}\to 0$. It is crucial, however, to keep the temperature $T$ above its critical value in order to avoid many-body effects (e.g. Bose-Einstein condensation). One may perform the vacuum limit for a fixed dimensionless $\frac{T}{T_{c}}$ such that the temperature goes to zero because $T_{c}$ scales $\sim k_{F}^{2}$.

Let us now examine the momentum-independent part of the atom inverse propagator $P_{\psi, k=0}(Q=0)=-\mu_{\psi}$, as well as its diatom counterpart $m_{\phi}^2\equiv P_{\varphi, k=0}(Q=0)$, in more detail\footnote{Flavor indices applicable for Systems II and III are suppressed in this section.}. For positive values, i. e. $\mu_{\psi}<0$, $m_{\varphi}^{2}>0$, they act as gaps for atoms and diatoms respectively. There is no Fermi surface in vacuum, hence $\mu_{\psi}\le0$. The system is above criticality in the vacuum limit, i.e. it is in the symmetric phase; hence $m_{\varphi}^{2}\ge 0$. These two conditions define a quadrant in the $m_{\varphi}^{2}-\mu_{\psi}$ plane. Moreover, due to the non-relativistic nature of the problem, the zero energy level can be shifted by an arbitrary constant. This is a result of the symmetries of our models. The real-time ($t=-i \tau$) version of the microscopic action $S=\Gamma_{\Lambda}$ in coordinate space ($t,\mathbf{x}$) is symmetric with respect to the energy shift symmetry \cite{Floerchinger}:
\beq \label{vac1a}
\psi\to e^{iE t}\psi \qquad \varphi\to e^{2iEt}\varphi \qquad \mu\to \mu +E.
\eeq
Since no anomaly of this symmetry is expected and our cutoff respects this symmetry (see below), this is a symmetry of the flow equations and the effective action $\Gamma_{k=0}$. Hence, by the appropriate energy shift, we can make one energy  state gapless, i.e. put it on the boundary of the quadrant in $m_{\varphi}^{2}-\mu_{\psi}$ plane. We end up with three distinct branches \cite{Diehl:2005ae}:
\begin{eqnarray}\label{VacCond}
  \begin{array}{l l l}
    { m_\varphi^2 >0, \quad \mu_{\psi} = 0  }& \text{atom phase}
      & (a^{-1} < 0) , \\
    { m_\varphi^2 = 0,\quad \mu_{\psi} < 0   }& \text{diatom phase}
      & (a^{-1} > 0) ,  \\
    { m_\varphi^2 = 0,\quad \mu_{\psi} = 0  }& \text{resonance}
      & (a^{-1} = 0).
  \end{array}
\end{eqnarray}
For System II the gapless state is the lowest energy state. In the atom phase $(a^{-1} < 0)$ diatoms $\varphi$ are gapped and the lowest excitation is an atom $\psi$. In the diatom phase $(a^{-1} > 0)$ the situation is reversed: $\varphi$ is the lowest excitation above the vacuum and $\psi$ has a gap $-\mu_{\psi}$, which can be interpreted as a half of the binding energy of $\varphi$, $\epsilon=2\mu_{\psi}$. At resonance $(a^{-1}=0)$ both, $\varphi$ and $\psi$, are gapless. 
For Systems I and III, and for small values of $|a|^{-1}$, one finds a whole spectrum of trions, bound states of three atoms, which have a lower energy then atoms and diatoms. This effect has been first predicted and calculated by Efimov in a quantum mechanical computation \cite{Efimov70}, and modifies the vacuum structure \cite{FSMW}. In the trion phase both atoms and diatoms show a gap, i.e. the ground state has $\mu_{\psi}<0$, $m_{\varphi}^{2}>0$. However, for an investigation of the excited Efimov states we may as well use the vacuum fixing condition (\ref{VacCond}). At resonance this corresponds to degenerate energy levels of the Efimov states and the atoms/diatoms, which becomes a good approximation for the high Efimov states which are close to the atom/diatom threshold \cite{FSMW}.

The vacuum limit, which we described above, leads to numerous mathematical simplifications. For example, all diagrams with loop lines pointing in the same direction \emph{vanish} in the vacuum limit. This can be demonstrated using the residue theorem for the frequency loop integration. Indeed, the inverse propagators have non-negative gaps and all considered diagrams have poles in the same half plane of the complex loop frequency. Thus we can close the contour such that it does not enclose any poles and the frequency integral vanishes. The argument works also for the 1PI vertices provided they have poles in the same half-plane as the propagators. This finding simplifies the RG analysis in vacuum considerably. For example, one can show that in vacuum the atom inverse propagator $P_{\psi}(Q)$ is not renormalized \cite{Diehl:2007th}. The only one-loop diagram, which renormalizes $P_{\psi}$, has inner lines pointing in the same direction, and therefore vanishes. It is sufficient to analyze only one-loop diagrams because the RG flow equation (\ref{M1}) has a general one-loop form \cite{Berges:2000ew}. Another very important simplification in vacuum comes from a special hierarchy, which is respected by the flow equations. We define the n-body sector as a set of 2n-point 1PI vertices written in terms of elementary atoms (in this sense $P_{\varphi}(Q)$ belongs to the two-body sector because $\varphi\sim \psi \psi$ is composed of two atoms). The vacuum hierarchy consists in the fact that the flow of the n-body sector \emph{is not influenced} by any higher-body sectors. The flow equations for the n-body sector simply decouple from the flow of the (n+1)-body sector (and higher). The observed hierarchy is a consequence of the diagrammatic simplification in vacuum. At finite density ($n\ne 0$) or temperature ($T\ne 0$) the decoupling of the low n vertices from the high n vertices is not valid anymore.

\section{Two-body sector: Exact Solution in the diatom phase for a positive scattering length}
\label{twobody}
The two-body sector truncation is defined by:
\beq \label{t1}
\Gamma_{k}=\Gamma_{k}(2)+\Gamma_{k}(3)
\eeq
in all three models ($\ref{M3}$),($\ref{M4}$) and ($\ref{M5}$). As mentioned in Sect. \ref{VacPro} the RG flows belonging to the two-body sector decouple from higher-body sectors in vacuum. Due to the non-renormalization of the atom propagator, it is sufficient to solve the flow equations only for the Yukawa coupling $h$ and the diatom inverse propagator $P_{\varphi}$.

It turns out that the Yukawa coupling is not renormalized in vacuum for all three models:
\beq \label{t2}
\partial_{t} h=0.
\eeq
Due to the U(1) phase symmetry, there is no one-loop Feynman diagram in our truncations (\ref{M3},\ref{M4},\ref{M5}), which renormalizes the Yukawa coupling $h$. The only one-loop diagram, which could contribute to the flow of $h$, contains the four-atom vertex $\lambda_{\psi}(Q_1, Q_2, Q_3)$. The vertex $\lambda_{\psi}(Q_1,Q_2,Q_3)$ is not renormalized in vacuum (see Appendix \ref{Gamma4}) and vanishes on all scales, provided its microscopic value is zero. The argument can be extended to a momentum-dependent Yukawa coupling $h(Q_1,Q_2,Q_3)$.

\begin{figure}[t]
\includegraphics[width=60mm]{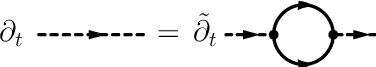}
\caption{\small{Schematic graphical representation of the flow for the inverse diatom propagator $P_{\varphi}$. Diatoms are denoted by dashed lines, atoms by solid lines.
}}
\label{fig:1}
\end{figure}

In order to solve the two-body sector, it remains to calculate the flow of the diatom inverse propagator $P_{\varphi}(Q)$, which is schematically shown in FIG. (\ref{fig:1}) and can be written as follows:
\beq\label{t3}
\partial_{t}P_{\varphi}(Q)=-\frac{2}{3+p}\int\limits_{L}\tilde{\partial}_{t}\frac{h^2}{(P_{\psi}(L)+R_{\psi}(L))(P_{\psi}(Q-L)+R_{\psi}(Q-L))},
\eeq
where $p=+1$ for bosons and $p=-1$ for fermions. It turns out that the flow of the inverse diatom propagator in System III is exactly the same as in the System II. In the last formula one has:
\beq \label{atomprop}
 P_{\psi}(Q)=(i\omega_{\mathbf{q}}+\mathbf{q}^2-\mu_{\psi})
\eeq
and $R_{\psi}(Q)$ stands for the atom regulator.

It is remarkable that using a special choice of the regulator, we can integrate the flow (\ref{t3}) exactly. We follow \cite{DKS, DGPW2} by choosing a regulator, which is frequency and momentum independent and has the form $R_{\psi}=k^2=\Lambda^{2}\exp{2t}$. This regulator has the advantage that it is Galilean invariant and hence the Galilean symmetry of the microscopic action is preserved during the RG evolution. First we perform the frequency loop integration in Eq. (\ref{t3}) with help of the residue theorem:
\begin{eqnarray} \label{t4}
\partial_{t}P_{\varphi}(Q)&=&-\frac{2}{3+p}\int \frac{d^{3}l}{(2\pi)^3}\tilde{\partial}_{t}\frac{h^2}{i \omega_{\mathbf{q}}+\mathbf{l}^2+(\mathbf{l}-\mathbf{q})^2-2\mu_{\psi}+2R_{\psi}}= \nonumber \\
&=&-\frac{2h^2}{3+p}\partial_{t}\int \frac{d^{3}l}{(2\pi)^3}\frac{1}{i\omega_{\mathbf{q}}+\mathbf{l}^2+(\mathbf{l}-\mathbf{q})^2-2\mu_{\psi}+2R_{\psi}},
\end{eqnarray}
where the second equality holds due to the non-renormalization of the Yukawa coupling $h$ and atom inverse propagator $P_{\psi}$, and thus $\tilde{\partial_t}\to\partial_t$.
Using the specific values of the regulator $R_{\psi}(t_{IR})=0$ and $R_{\psi}(t=0)= \Lambda^{2}$ we integrate out the flow equation from the UV scale $t=0$ to the IR scale $t_{IR}=-\infty$ and obtain:
\begin{eqnarray} \label{t5}
P^{IR}_{\varphi}(Q)-P^{UV}_{\varphi}(Q)&=&-\frac{2h^2}{3+p}\int \frac{d^{3}l}{(2\pi)^{3}}\left( \frac{1}{i\omega_{\mathbf{q}}+\mathbf{l}^2+(\mathbf{l}-\mathbf{q})^2-2\mu_{\psi}}-\frac{1}{i\omega_{\mathbf{q}}+\mathbf{l}^2+(\mathbf{l}-\mathbf{q})^2-2\mu_{\psi}+2\Lambda^{2}} \right) \nonumber \\
&=&-\frac{h^2}{3+p}\int\frac{dl}{2\pi^2}\left( \frac{l^2}{l^2+\big(\frac{i\omega_{\mathbf{q}}}{2}+\frac{\mathbf{q}^2}{4}-\mu_{\psi}\big)} -\frac{l^2}{l^2+\big(\frac{i\omega_{\mathbf{q}}}{2}+\frac{\mathbf{q}^2}{4}-\mu_{\psi}+\Lambda^{2}\big)} \right) \nonumber \\
&=&-\frac{h^2}{4\pi (3+p)}\big(\Lambda-\sqrt{\frac{i\omega_{\mathbf{q}}}{2}+\frac{\mathbf{q}^2}{4}-\mu_{\psi}} \big).
\end{eqnarray}
The last identity assumes $\Lambda>> |\mu_{\psi}|, |\mathbf{q}|, |\omega_{\mathbf{q}}|$.

At this point we must fix the initial condition $P^{UV}_{\varphi}(Q)$ at $k=\Lambda$ in order to obtain the physical inverse propagator $P^{IR}_{\varphi}(Q)$ at $k=0$. This is done in \cite{Diehl:2005ae,Diehl:2007th,DKS} and we follow the same steps here. For broad resonances with $h^2\to\infty$ the inverse diatom propagator at the microscopic scale $\Lambda$ is given by:
\beq \label{t6}
P^{UV}_{\varphi}=\nu(B)+\delta\nu, \qquad \nu(B)=\mu_{B}(B-B_{0}).
\eeq
Here $\nu(B)$ is the detuning of the magnetic field $B$ which measures the distance to the Feshbach resonance located at $B_{0}$. The  magnetic moment of the diatom is denoted by $\mu_{B}$. The counter term $\delta\nu$ depends on the ultraviolet cutoff $\Lambda$. Neglecting a possible background scattering length $a_{bg}$, the scattering length $a$ and the detuning $\nu (B)$ are related by \cite{Diehl:2005ae}:
\beq \label{t7}
a=-\frac{h^2}{4\pi (3+p)\nu(B)}.
\eeq
Thus the Yukawa coupling is proportional to the square root of the width of the Feshbach resonance. For narrow Feshbach resonances ($h\to 0$) perturbation theory is applicable, while for broad Feshbach resonances ($h\to \infty$), which are of main interest in our work, the problem becomes strongly coupled. Using Eq. (\ref{t6}) and (\ref{t7}), we rewrite Eq. (\ref{t5}) as:
\beq \label{t8}
P^{IR}_{\varphi}(Q)-\delta\nu+\frac{h^2}{4\pi(3+p)a}=-\frac{h^2}{4\pi (3+p)}\big(\Lambda-\sqrt{\frac{i\omega_{\mathbf{q}}}{2}+\frac{\mathbf{q}^2}{4}-\mu_{\psi}} \big).
\eeq   
At this point the momentum independent counter term $\delta\nu$ can be identified:
\beq \label{t9}
\delta\nu=\frac{h^2}{4\pi (3+p)}\Lambda
\eeq
and we obtain our final result for the $k$-dependent inverse diatom propagator $P_{\varphi, k}(Q)$:
\beq \label{t9a}
P_{\varphi, k}(Q)=\frac{h^2}{4\pi(3+p)}\Big(-a^{-1}+\sqrt{\frac{i\omega_{\mathbf{q}}}{2}+\frac{\mathbf{q}^2}{4}-\mu_{\psi}+k^{2}} \Big).
\eeq
The wave-function renormalization $Z_{\varphi, k}$ can now be defined:
\beq \label{t9b}
Z_{\varphi, k}\equiv \frac{\partial P_{\varphi, k}(Q)}{\partial (i \omega_{\mathbf{q}})}{\bigg|}_{\omega_{\mathbf{q}}=0}=\underbrace{\frac{h^{2}}{4\pi(3+p)}}_{\tilde{Z}} \frac{1}{4\sqrt{k^{2}-\mu_{\psi}}},
\eeq
and the IR inverse diatom propagator $P_{\varphi}(Q)$ reads:
\beq \label{t10}
P_{\varphi}(Q)\equiv P_{\varphi, k=0}(Q)=\frac{h^2}{4\pi(3+p)}\Big(-a^{-1}+\sqrt{\frac{i\omega_{\mathbf{q}}}{2}+\frac{\mathbf{q}^2}{4}-\mu_{\psi}} \Big).
\eeq

In vacuum and for positive scattering length ($a>0$) the vacuum condition, (\ref{VacCond}), $P_{\varphi}(Q=0)=m_{\varphi}^{2}=0$, must be fulfilled. This leads to:
\beq \label{t11}
a=\frac{1}{\sqrt{-\mu_{\psi}}}.
\eeq  
For positive scattering lengths in vacuum $-\mu_{\psi}$ is a positive gap of the atom $\psi$ and can be interpreted as half of the binding energy of the diatom $\epsilon_{\varphi}$. Hence, the binding energy can be expressed as:
\beq \label{t12}
\epsilon_{\varphi}=2\mu_{\psi}=-\frac{2}{a^{2}}, \qquad \epsilon_{\varphi}=-\frac{1}{M a^{2}}.
\eeq
The second equation is expressed in conventional units and is the well-known universal relation for the binding energy of the shallow diatom \cite{Braaten04}. It should be mentioned here that the two-body sector can also be solved exactly using a non-relativistic version of the Litim cutoff \cite{Birse}, which is optimized in the sense of \cite{Litim,Pawlowski:2005xe}. The drawback of this cutoff is that it breaks Galilean symmetry and one has to put some Galilean non-invariant counter terms into $P^{UV}_{\varphi}(Q)$ to restore Galilean symmetry in the IR.

It is important to stress the appearance of universality in the broad resonance limit ($h^{2}\to\infty$) \cite{Diehl:2005ae}: The IR physics becomes insensitive to the initial conditions in the UV. For example, one may consider possible momentum-dependent  modifications of the microscopic inverse propagator $P_{\varphi, k=\Lambda}$, which result in deviations from an exactly pointlike form. Their effect on $P_{\varphi, k=0}$ is suppressed by $h^{-2}$ with respect to the quantum loop contribution and it therefore becomes irrelevant in the broad resonance limit. In vacuum, and for $h\to\infty$, the only physically relevant scale is given by the scattering length $a$.\\

Physics becomes \emph{completely universal} if we perform the unitarity limit, $h^{2}\to\infty$ (broad resonance limit) and $a^{-1}\to 0$ (resonance limit) \cite{Ho}. In vacuum, all scales drop out in this limit. The atom and diatom inverse propagators take the following form:
\beq \label{t13}
P_{\psi}(Q)=i\omega_{\mathbf{q}}+\mathbf{q}^2, \qquad P_{\varphi}(Q)=\tilde{Z}\sqrt{\frac{i\omega_{\mathbf{q}}}{2}+\frac{\mathbf{q}^2}{4}}.
\eeq  
An alternative quantum-mechanical derivation of Eq. (\ref{t13}) can be found in Appendix \ref{separable}.

Let us perform a scaling dimension counting in the unitary limit\footnote{We denote a scaling dimension of some quantity $X$ by $[X]$.}. We start with the fact that $[\Gamma]=0$. In non-relativistic physics, energy scales as two powers of momentum\footnote{This is known as the dynamical exponent $z=2$.} and the free field scaling reads:
\beq \label{t14}
[q]=1, \qquad [\omega]=2, \qquad [\psi]=3/2, \qquad [\varphi]=3/2, \qquad [h]=1/2.
\eeq
For the universal interacting theory the scaling of $\varphi$ is modified according to Eq. (\ref{t13}). The scaling dimension of the diatom field $\varphi$ and Yukawa coupling $h$ can be computed from the Yukawa term and the kinetic term of $\varphi$:
\begin{equation} \label{t15}
[h]+[\varphi]+2[\psi]=5 \qquad \text{Yukawa}, \qquad
2[h]+2[\varphi]+1=5 \qquad \text{Kinetic}.
\end{equation}
This system is degenerate and we get a solution $[h]=\alpha$ and $[\varphi]=2-\alpha$, where $\alpha$ is some real number. The absence of a scaling of $h$ in Eq. (\ref{t2}), however, fixes $[h]=0$ and $[\varphi]=2$. Note that the scaling of the diatom field at unitarity is different to the scaling of the atom field $\psi$. This is a manifestation of the fact that the scaling in the two-body sector is governed by a fixed point\footnote{called a unitarity fixed point}, which is different from the Gaussian fixed point \cite{Sachdev06, DGPW2}. Exactly at unitarity no obvious scales are left in the problem and the theory seems to be scale invariant. Even more, at the two-body sector level the theory seems to be an example of a non-relativistic conformal field theory (NRCFT)\footnote{Another example of NRCFT in two spatial dimensions is a theory of anyons \cite{Nishida}.}. This type of theories are symmetric with respect to the \emph{Schr\"odinger group}, which is an extension of the Galilean symmetry group (for more details see Appendix \ref{NRCFT}). It is known, however, that the Schr\"odinger symmetry can be broken by a quantum anomaly in higher-body sectors \cite{Camblong}. The fate of the Schr\"odinger symmetry is different for the different systems considered. For the resonantly interacting particles (Systems I and III) it was demonstrated by Efimov \cite{Efimov70} a long time ago  that in the three-body sector the continuous scaling symmetry, which is a part of the Schr\"odinger symmetry, is broken to the discrete scaling subgroup $Z$ \cite{Braaten04}. This manifests itself in the appearance of a geometric spectrum of bound states in the three body sector, which is called the Efimov effect. For the System II of SU(2) symmetric fermions it is believed that the Schr\"odinger symmetry is not broken in the higher sectors of the theory and that this is a real example of an NRCFT \cite{Nishida}.

To summarize, in this section we have solved exactly the two-body sector in vacuum for a positive scattering length. The solution (\ref{t10}) was obtained for the specific initial conditions $h_{k=\Lambda}(Q_1,Q_2)=h$, $\lambda_{\psi, k=\Lambda}=0$ and $P_{\varphi, k=\Lambda}(Q)$ given by Eq. (\ref{t6}). This choice corresponds to a pointlike microscopic atom interaction. However, the presented calculations can be generalized to an arbitrary boson mediated atom interaction with $\lambda_{\psi, k=\Lambda}=0$ while $h_{k=\Lambda}(Q_1,Q_2)$ and $P_{\varphi, k=\Lambda}(Q)$ can be chosen freely.


\section{Three-Body Sector: Flow equations}
\label{Threeflow}
The main emphasis of this work is devoted to the analysis of the three-body sector of the three models (\ref{M3}), (\ref{M4}) and (\ref{M5}) in the unitarity limit. We demonstrate the appearance of the Efimov effect in Systems I and III and its absence in System II from the field theoretical RG perspective. In the present section we formulate a flow equation for the coupling $\lambda_{3}(Q_{1},Q_{2},Q_{3})$ and make some general simplifications. In the next section we use the pointlike approximation for $\lambda_{3}(Q_{1},Q_{2},Q_{3})$ to solve the problem. 
The last two sections are devoted to the solution of the general momentum-dependent form of the flow equation.

The closed, exact solution for the two-body sector provides a simple strategy for a computation of the coupling $\lambda_3$ in the three-body sector. In general, one may introduce separate cutoffs $R_{\psi}$ and $R_{\varphi}$ for the atoms $\psi$ and diatoms $\varphi$. The presence of the cutoff $R_{\varphi}$ does not affect our computation in the two-body sector. We may therefore first lower the cutoff $R_{\psi}$ from $\Lambda^{2}$ to zero, while keeping $R_{\varphi}$ fixed, and subsequently lower $R_{\varphi}$ to zero in a second step \cite{Berges:2000ew}. As the result of the first step the diatom inverse propagator $P_{\varphi}$ is modified according to Eq. (\ref{t10}). This step also induces diatom interactions, as for example a term $\sim (\varphi^{*}\varphi)^2$. However, these interactions belong to the four-body and higher sectors. By virtue of the vacuum hierachy, they do not influence the flow of $\lambda_3$. For the second step of our computation we can therefore use a version of the flow equation where only the diatom cutoff $R_{\varphi}$ is present. In this flow equation $P_{\varphi}$ and $P_{\psi}$ are fixed according to Eqs. (\ref{t10}) and (\ref{atomprop}).

For the diatoms we use a sharp cutoff:
\beq \label{th3a}
R_{\varphi}(Q,k)=P_{\varphi}(Q)\left(\frac{1}{\theta(|q|-k)}-1 \right).
\eeq
The special feature of this cutoff is that the regularized diatom propagator takes a simple form:
\beq \label{th3b}
\frac{1}{P_{\varphi}(Q)+R_{\varphi}(Q,k)}=\theta(|q|-k)\frac{1}{P_{\varphi}(Q)}.
\eeq
Thus the propagator is cut off sharply at the sliding scale $k$. Our choice of the cutoff is motivated by technical simplicity as well as effective theory \cite{Braaten04} and quantum mechanical \cite{Efimov70} approaches to this problem. The advantage of this cutoff is the property of locality in the momentum space, which means that it chops off momentum shells locally. In the three-body sector we are interested not only in the IR value of the atom-diatom vertex $\lambda_{3}$, but also in the flow at all scales. 

Let us now calculate the flow equation of the 1PI atom-diatom vertex $\lambda_{3}$. For SU(3) fermions there are two atom-diatom vertices, $\lambda_{3a}$ and $\lambda_{3b}$, and we postpone the analysis of this model to the end of the section. In Minkowski space (the real time version of our theory) the atom-diatom scattering amplitude is given by the amputated connected part of the Green's function $\langle 0 | \varphi\psi \varphi^\dagger\psi^\dagger |0 \rangle$, and thus it can be simply calculated from the knowledge of $\lambda_{3}$. We first consider the kinematics of the problem. The 1PI atom-diatom vertex $\lambda_{3}(Q_1,Q_2,Q_3)$ depends generally on three four vectors, i.e. six independent rotation invariant variables in the center-of-mass frame.  We take the incoming atom and diatom to have momenta $\mathbf{q}_{1}$ and $-\mathbf{q}_{1}$, and energies $E_{\psi1}$ and $E-E_{\psi1}$, while the outgoing atom and diatom have momenta $\mathbf{q}_{2}$ and $-\mathbf{q}_{2}$ and energies $E_{\psi2}$ and $E-E_{\psi2}$. We denote the vertex in the center-of-mass frame by $\lambda_{3}(Q_{1}^{\psi},Q_{2}^{\psi},E)$ (see FIG 2.). This configuration is in general off-shell which is necessary since, in the flow equations, the vertex also appears inside a loop.    
\begin{figure}[t]
\includegraphics[width=100pt, height=100pt]{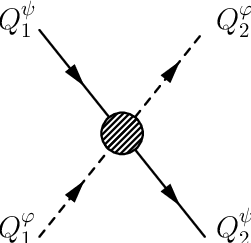}
\caption{\small{Kinematics of the vertex $\lambda_{3}(Q_{1}^{\psi},Q_{2}^{\psi},E)$ in the center-of-mass frame. The atoms and diatoms have momenta $Q_{1}^{\psi}=(E_{\psi 1},\mathbf{q}_{1})$ , $Q_{1}^{\varphi}=(-E_{\psi 1}+E,-\mathbf{q}_{1})$ and $Q_{2}^{\psi}=(E_{\psi 2},\mathbf{q}_{2})$, $Q_{2}^{\varphi}=(-E_{\psi 2}+E,-\mathbf{q}_{2})$.
}}
\label{fig:2}
\end{figure}

\begin{figure}[t]
\includegraphics[width=120mm]{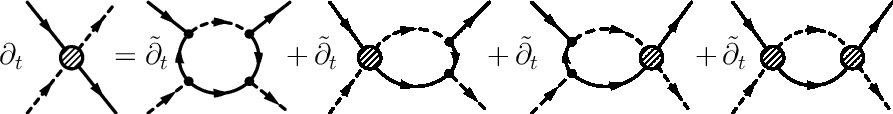}
\caption{\small{Graphical representation of the flow equation for $\lambda_{3}$. Full lines denote atoms $\psi$ and dashed lines diatoms $\varphi$. The shaded circle denotes $\lambda_{3}$.
}}
\label{fig:3}
\end{figure}

In Minkowski space\footnote{The flow equation of the effective action (\ref{M1}) is formulated in Euclidean spacetime (imaginary time formalism). In order to obtain the flow equation in Minkowski space, it is sufficient to take external frequencies $\omega_{ext}$ to be imaginary, i.e. perform a transformation $\omega_{\text{ext}}\to i\omega_{\text{ext}}$, which is the inverse Wick rotation. } the flow equation for the atom-diatom vertex $\lambda_{3}$ for the Systems I and II reads:
\begin{eqnarray}\label{th4}
\partial_t \lambda_{3}(Q^{\psi}_1,Q^{\psi}_2; E) &=&   \int\limits_L \tilde{\partial_t}\frac{\theta(|\mathbf{l}|-k)}{P_{\psi}(L)  P_{\varphi}(-L+Q)} \Big[C \lambda_{3}(Q_1^{\psi},L;E) \lambda_{3}(L,Q_2^{\psi} ;E) \\\nonumber
&&\qquad \qquad + \frac{B}{2}\big(  \frac{h^{2}}{P_{\psi}(-L + Q_1^{\varphi})} \lambda_{3}(L,Q_2^{\psi};E) 
 +  \lambda_{3}(Q_1^{\psi},L;E) \frac{h^2}{P_{\psi}(-L + Q_2^{\varphi})}\big)\\\nonumber
&&\qquad\qquad + A\frac{h^2}{P_{\psi}(-L + Q_1^{\varphi})}\,\frac{h^2}{P_{\psi}(-L + Q_2^{\varphi})}\Big],
\end{eqnarray}
where $Q=Q_1^{\varphi}+Q_1^{\psi}=(E,\mathbf{0})$. The coefficients $A$, $B$ and $C$ for Systems I, II can be found in TABLE I. The graphical representation of this equation is depicted in FIG. {\ref{fig:3}}. The scale derivative on the RHS acts only on the cutoff and can be computed easily, $\tilde{\partial_{t}}\theta(|\mathbf{l}|-k)=-k\delta(|\mathbf{l}|-k)$.
\begin{table}
\begin{center}
\begin{tabular}{|c|c|c|c|}
\hline
Model & A & B & C \\
\hline
System I & 1 & 2 & 1 \\ 
\hline
System II & 1 & -2 & 1 \\
\hline
System III(a) & 1 & -2 & 1 \\
\hline
System III(b) &4 & 4 & 1 \\
\hline
\end{tabular}
\end{center}
\caption{Numerical coefficients $A$, $B$ and $C$ in the flow equation (\ref{th4}) for the three examined systems. In the case of System III(a) we consider the scattering of the type $\psi_{i}\varphi_{j}\to\psi_{i}\varphi_{j}$  with $i\ne j$, while System III(b) corresponds to the 
the vertex $\lambda_3=3\lambda_{3a}+\lambda_{3b}$}
\end{table}

Fortunately, the flow equation can be simplified considerably. First note that there is only one inverse propagator $P_{\psi}(L)$ with a loop momentum $L$ of positive sign in Eq. (\ref{th4}). For this reason the whole integrand in Eq. $(\ref{th4})$ has a single frequency pole in the upper half plane. Thus the frequency integration in Eq. $(\ref{th4})$ can be performed with the help of the residue theorem by performing the substitution $\omega_{\mathbf{l}}\to i \mathbf{l}^2$. This puts the atom in the loop on-shell, corresponding to $P_\psi(L)$ in Eq. (\ref{th4}). We obtain a simpler equation if we also put the energies of the incoming and outgoing atoms on-shell ($Q_{1}^{\psi}=(i\mathbf{q}_{1}^2, \mathbf{q}_{1})$, $Q_{2}^{\psi}=(i\mathbf{q}_{2}^2, \mathbf{q}_{2})$). The diatoms in the loop in Eq. (\ref{th4}) are generally off-shell. To solve this ``half-off-shell'' equation only the values $\lambda_{3}(\mathbf{q}_{1},\mathbf{q}_{2},E)\equiv\lambda_{3}(Q_{1}^{\psi}=(i\mathbf{q}_{1}^2, \mathbf{q}_{1}),Q_{2}^{\psi}=(i\mathbf{q}_{2}^2, \mathbf{q}_{2}),E)$ are needed \cite{Braaten04}.

Our aim is the calculation of the atom-diatom scattering amplitude at \emph{low energies and momenta}. For low momenta the dominant contribution is given by s-wave scattering. In principle, the right hand side of Eq. (\ref{th4}) has also contributions from higher partial waves, which we neglect in our approximation and simplify the flow equation (\ref{th4}) by projecting on the s-wave. This is done by averaging Eq. (\ref{th4}) over the cosine of the angle between incoming momentum $\mathbf{q}_{1}$ and outgoing momentum $\mathbf{q}_{2}$. Introducing the averaged 1PI renormalized vertex, which depends on three scalar variables:
\beq \label{th5}
\lambda_{3}(q_{1},q_{2},E)\equiv\frac{1}{2h^2}\int^{1}_{-1}d(\cos\theta)\lambda_{3}(\mathbf{q}_{1},\mathbf{q}_{2},E),
\eeq
we end up with the flow equation:
\begin{eqnarray} \label{th6}
\partial_{t}\lambda_{3}(q_1,q_2,E)&=&-\frac{2(3+p)}{\pi}\frac{k^3}{\sqrt{\frac{3k^2}{4}-\frac{E}{2}-i\epsilon}} \left[ C \lambda_{3}(q_{1},k,E)\lambda_{3}(k,q_{2},E)+ \right. \nonumber \\
&&\left. \frac{B}{2}\left\{ \lambda_{3}(q_{1},k,E)G(k,q_{2})+G(q_{1},k)\lambda_{3}(k,q_{2},E) \right\}+A G(q_{1},k)G(k,q_{2}) \right],
\end{eqnarray}
where the symmetric function $G(q_{1},q_{2})$ is defined by\footnote{This is in fact a s-wave projected tree (one-particle-reducible) contribution to the fully connected atom-diatom vertex $\lambda_{3}(q_1,q_2,E)$.}:
\beq \label{th7}
G(q_{1},q_{2})=\frac{1}{4q_{1}q_{2}}\log\frac{q_{1}^2+q_{2}^2+q_{1}q_{2}-\frac{E}{2}-i\epsilon}{q_{1}^2+q_{2}^2-q_{1}q_{2}-\frac{E}{2}-i\epsilon}.
\eeq
The infinitesimally positive $i\epsilon$ term arises from the Wick rotation and makes both Eq. (\ref{th6}) and (\ref{th7}) well-defined. It is remarkable that Eq. (\ref{th6}) is completely independent of the Yukawa coupling $h$ and thus is a well-defined equation in the limit of infinite $h$.

For $SU(3)$ fermions the situation is more complicated because there are two vertices $\lambda_{3a}$ and $\lambda_{3b}$ in our truncation (\ref{M5}). To extract the flow equation for $\lambda_{3a}$ we consider the scattering channel $\varphi_{i}\psi_{i}\to \varphi_{j}\psi_{j}$ with $i \ne j$ (e.g. $\varphi_{1}\psi_{1}\to \varphi_{2}\psi_{2}$). After performing the same steps as for System I and II, we end up with a flow equation:
\begin{eqnarray} \label{th8}
\partial_{t}\lambda_{3a}(q_1,q_2,E)&=&-\frac{2(3+p)}{\pi}\frac{k^3}{\sqrt{\frac{3k^2}{4}-\frac{E}{2}-i\epsilon}} \left[ 3 \lambda_{3a}(q_{1},k,E)\lambda_{3a}(k,q_{2},E)+2\lambda_{3a}(q_{1},k,E)\lambda_{3b}(k,q_{2},E)+ \right. \nonumber \\
&&\left. 2\left\{ \lambda_{3a}(q_{1},k,E)G(k,q_{2})+G(q_{1},k)\lambda_{3a}(k,q_{2},E) \right\} + \right. \nonumber \\
&&\left. +\left\{ \lambda_{3b}(q_{1},k,E)G(k,q_{2})+G(q_{1},k)\lambda_{3b}(k,q_{2},E) \right\}
 +G(q_{1},k)G(k,q_{2}) \right], 
\end{eqnarray}
where $p=-1$ and $G(q_1,q_2)$ is defined in Eq. (\ref{th7}). Note, that the coupling $\lambda_{3b}$ appears in the flow equation for $\lambda_{3a}$. The flow equation for $\lambda_{3b}$ can be extracted by considering the scattering channel $\varphi_{i}\psi_{j}\to \varphi_{i}\psi_{j}$ with $i \ne j$ (e.g. $\varphi_{2}\psi_{1}\to \varphi_{2}\psi_{1}$):
\begin{eqnarray} \label{th9}
\partial_{t}\lambda_{3b}(q_1,q_2,E)&=&-\frac{2(3+p)}{\pi}\frac{k^3}{\sqrt{\frac{3k^2}{4}-\frac{E}{2}-i\epsilon}} \left[ \lambda_{3b}(q_{1},k,E)\lambda_{3b}(k,q_{2},E)+ \right. \nonumber \\
&&\left. -\left\{ \lambda_{3b}(q_{1},k,E)G(k,q_{2})+G(q_{1},k)\lambda_{3b}(k,q_{2},E) \right\}+G(q_{1},k)G(k,q_{2}) \right]. 
\end{eqnarray}
This equation is completely decoupled from Eq. (\ref{th8}) and has exactly the same form as Eq. (\ref{th6}) for $SU(2)$ fermions. The reason for this is simple: The RG equation (\ref{th9}) has the graphical representation depicted in FIG. \ref{fig:3}. It turns out that in this channel only one type of diatom (in our example $\phi_2$) and two types of atoms ($\psi_1$ and $\psi_3$) appear, which is exactly the same as in the case of $SU(2)$ fermions. Remarkably, it is possible to introduce a linear combination $\lambda_{3}\equiv3\lambda_{3a}+\lambda_{3b}$ for $SU(3)$ fermions, which has a simple flow equation of the form (\ref{th6}) with coefficients $A$, $B$ and $C$ given in TABLE I (forth line). We call this System III(b). In the SU(3) fermion model the diatom-atom in-state  $\varphi_{i}\psi_{i}$ can lead to the different diatom-atom out-states $\varphi_{1}\psi_{1}$, $\varphi_{2}\psi_{2}$ and $\varphi_{3}\psi_{3}$. If the diatom-atom out-state is not a final but only an intermediate state(e.g. one is interested in the scattering into a three atom final state), we must sum the scattering amplitudes for all possible atom-diatom pairs. It easy to show that $\lambda_{3}=3\lambda_{3a}+\lambda_{3b}$ corresponds to the 1PI contribution to the full scattering amplitude $\varphi_{i}\psi_{i}\to anything$ (e.g. $\varphi_{i}\psi_{i}\to \varphi_{1}\psi_{1}+\varphi_{2}\psi_{2}+\varphi_{3}\psi_{3}$).\\ 

To summarize, although at first sight it seems that for SU(3) fermions we must solve a system of two flow equations, it turns out that for the two specific situations it is sufficient to solve only \emph{one equation} (\ref{th6}). This equation is the main result of this section. In the next sections we solve this final version of the RG flow equation for atom-diatom 1PI vertex for all three systems using various approaches.

\section{Three-body sector: Pointlike approximation}
\label{triatom}

In this section the flow equation (\ref{th6}) will be solved employing a simple and intuitive pointlike approximation. The 1PI vertex $\lambda_{3}(q_1,q_2,E)$ will be replaced by a single momentum-independent coupling $\lambda_{3}(E)$.
In the low energy limit ($E\to 0$) the flow equation (\ref{th6}) takes a simple form in the pointlike approximation:
\beq \label{ta1}
\partial_{t}\lambda_{3}^{R}=-\frac{4(3+p)}{\sqrt{3}\pi}\left[ \frac{A}{4}+\frac{B}{2}\lambda_{3}^{R}+C(\lambda_{3}^{R})^{2} \right]+2\lambda_{3}^{R},
\eeq
where we use $G(q\to 0,k)\to\frac{1}{2k^2}$ from Eq. (\ref{th7}). The renormalized coupling is defined as $\lambda_{3}^{R}=\lambda_{3}k^{2}$. This definition is motivated by a simple power counting near the unitary fixed point ($[\varphi\psi\varphi^{*}\psi^{*}]=7\to [\lambda_{3}]=-2\to [\lambda_{3}^{R}]=0$). The RHS of Eq. (\ref{ta1}) is a quadratic polynomial in $\lambda_{3}^{R}$ with constant coefficients. This type of equation is discussed in Appendix \ref{triatomap}. The behavior of the solution is governed by the \emph{sign} of the discriminant $D$ of the quadratic polynomial on the RHS of Eq. (\ref{ta1}):
\begin{itemize}
\item $D>0$ -- fixed point solution
\item $D=0$ -- see Appendix \ref{triatomap}
\item $D<0$ -- periodic limit cycle solutions with a period $T=\frac{2\pi}{\sqrt{-D}}$.
\end{itemize} 

The discriminant is given by:

\beq \label{ta1a}
D=4\left(1-\frac{B(3+p)}{\sqrt{3}\pi} \right)^{2}-\frac{16AC(3+p)^2}{3\pi^2}.
\eeq

\begin{table}
\begin{center}
\begin{tabular}{|c|c|c|c|}
\hline
Model & $D$ & $T$ & $s_{0}$ \\
\hline
System I & -7.762 & 2.255 & 1.393 \\ 
\hline
System II & 9.881 & -- & -- \\
\hline
System III(a) & 9.881 & -- & -- \\
\hline
System III(b) & -7.762 & 2.255  & 1.393 \\
\hline
\end{tabular}
\end{center}
\caption{Discriminant $D$, temporal RG period $T$ (if applicable) and Efimov parameter $s_{0}$ (if applicable) in the pointlike approximation for Systems I, II, III(a) and III(b).}
\end{table}

In the special case of the Systems I, II and III the solution in the pointlike approximation is summarized in TABLE II. For Systems II and III(a) we find the solution with a fixed point with vanishing anomalous dimension $\eta=0$ in the IR (see Appendix \ref{triatomap}). For Systems I and III(b) the situation is completely different. We obtain a periodic limit cycle solution of the form $\lambda_{3}^{R}(t)\sim \tan \left[T t \right]$. The intuitive interpretation of this solution is that during the RG flow we hit three-body diatom-atom bound states, which manifest themselves as divergences of $\lambda_{3}^{R}$. In the unitary limit there are infinitely many of these bound states, which are equidistant in a logarithmic scale. The continuous scaling symmetry is broken to the discrete $Z$ group. This is the well-known Efimov effect \cite{Efimov70, Braaten04}, which indeed is present for equivalent bosons (System I) and is absent in the case of SU(2) fermions. In the case of equivalent bosons (System I) the Efimov result is:
\beq \label{ta2}
\frac{E_{n+1}}{E_{n}}=\exp(-2\pi/s_{0})
\eeq   
with $E_{n+1}$ and $E_{n}$ denoting neighboring bound state energies. The Efimov parameter $s_{0}$ is given by the solution of a transcendental equation and one finds $s_{0}\approx 1.006$\cite{Efimov70}. By dimensional arguments we can connect the artificial sliding scale $k^{2}$ with the scattering energy $E$ as $E\sim k^2$ \cite{FSMW}. The proportionality factor disappears in the ratio of the energies and hence the Efimov parameter can be read off from the RG period:
\beq \label{ta3}
\frac{k^{2}_{n+1}}{k^{2}_{n}}=\frac{E_{n+1}}{E_{n}}=\exp(-2T) \Rightarrow s_{0}=\frac{\pi}{T}.
\eeq
The values of the Efimov parameter for Systems I and III(b) can be found in TABLE II. We obtain $s_{0}\sim 1.393$, which differs from the correct result by $40\%$. In the next two sections we demonstrate that the simple pointlike approximation is too crude to get the correct quantitative agreement. Nevertheless it provides us with the first hint how the Efimov effect appears also in the functional renormalization group framework.

\section{Three-body sector: Systems I and II}
\label{one-two}
In this section we only discuss Systems I and II leaving the analysis of System III to the next section. It turns out that in these two cases the flow equation (\ref{th6}) for $E=0$ can  be formally solved exactly. For two-component fermions this was shown by Diehl et. al. in \cite{DKS}. To find the exact solution most easily we perform the following redefinition:
\beq \label{bf1}
f_{t}(t_{1},t_{2},E)\equiv4(3+p)q_{1}q_{2}\lambda_{3}(q_{1},q_{2},E) \qquad g(t_{1},t_{2})\equiv4(3+p)q_{1}q_{2}G(q_{1},q_{2}), 
\eeq
where, from now on, we prefer to work with logarithms of momenta $t_{1}=\ln(q_{1}/\Lambda)$ and $t_{2}=\ln(q_{2}/\Lambda)$. As before $p=+1$  ($p=-1$) in the case of bosons (fermions). The RG scale dependence of the reduced atom-diatom vertex $f_{t}(t_1,t_2,E)$ is denoted by the subscript $t$. It is important to stress that we are generally interested in the solution of Eq. (\ref{th6}) for the scattering of particles of non-zero energy $E$. Nevertheless, we observe that the energy $E$ cuts off the RG flow in Eq. (\ref{th6}) in a similar way as the regulator (\ref{th3a}). With this relation between $k^2$ and $E$ in mind, the coupling for $k\ne 0$ and $E=0$ imitates the effect of a non-zero energy of the scattering particles, i.e. $k=0$, $E\ne0$.

 The flow equation at vanishing energy $E=0$ now reads:
\beq \label{bf2}
\partial_{t}f_{t}(t_{1},t_{2})=-\frac{1}{\sqrt{3}\pi}\left[ A g(t_{1},t)g(t,t_{2})+\frac{B}{2}\left\{f_{t}(t_{1},t)g(t,t_{2})+g(t_{1},t)f_{t}(t,t_{2}) \right\}+C f_{t}(t_{1},t) f_{t}(t,t_{2}) \right].
\eeq 
We assume that in the UV the reduced atom-diatom 1PI vertex is vanishing, i.e. the initial condition is $f_{t=0}(t_{1},t_{2})=0$. In general we are dealing with the Riccati differential equation in matrix form, where both matrices $g$ and $f_{t}$ have a continuous index running in the interval $t_{1},t_{2}\in(-\infty,0)$. The RHS of Eq. (\ref{bf2}) is a complete square, which is a special feature of the Systems I and II (see TABLE I). In order to find the formal solution of Eq. (\ref{bf2}) we define:
\beq \label{bf3}
\bar{f}_{t}(t_{1},t_{2})=p f_{t}(t_{1},t_{2})+g(t_{1},t_{2}),
\eeq
which can be recognized as the reduced, fully connected atom-diatom vertex.  
The flow equation for the full vertex $\bar{f}_{t}(t_{1},t_{2})$ with the initial condition takes the simple form:
\beq \label{bf4}
\partial_{t}\bar{f}_{t}(t_{1},t_{2})=-\frac{p}{\sqrt{3}\pi}\bar{f}_{t}(t_{1},t)\bar{f}_{t}(t,t_{2}), \qquad \bar{f}_{t=0}(t_{1},t_{2})=g(t_{1},t_{2}).
\eeq
It is convenient to rewrite Eq. (\ref{bf4}) in matrix notation ($\bar{f}_{t}(t_{1},t_{2})\to \bar{f}_{t}$):
\beq \label{bf5}
\partial_{t}\bar{f}_{t}=-\frac{p}{\sqrt{3}\pi}\bar{f}_{t}\cdot A_{t} \cdot \bar{f}_{t}, \qquad \bar{f}_{t=0}=g,
\eeq
where $A_{t}$ has matrix elements $A_{t}(t_{1},t_{2})=\delta(t-t_{1})\delta(t-t_{2})$ and matrix multiplication denotes t-integration. Multiplying both sides of Eq. (\ref{bf5}) from the left and right by $\bar{f}_{t}^{-1}$ we obtain:
\beq \label{bf6}
\partial_{t}\bar{f}_{t}^{-1}=-\bar{f}_{t}^{-1}\cdot \partial_{t}\bar{f}_{t} \cdot \bar{f}_{t}^{-1} =\frac{p}{\sqrt{3}\pi} A_{t}, \qquad \bar{f}_{t=0}^{-1}=g^{-1},
\eeq
which is formally solved by:
\beq \label{bf7}
\bar{f}_{t}=\left(I+\frac{p}{\sqrt{3}\pi}\int_{0}^{t} ds g\cdot A_{s} \right)^{-1}\cdot g
\eeq
for $t\in(-\infty,0)$. $I$ denotes the identity matrix.

In the IR limit $t\to -\infty$, which corresponds to integration of all quantum fluctuations, $\bar{f}\equiv \bar{f}_{t=-\infty}$ solves the following matrix equation:
\beq \label{bf8}
\bar{f}=g+\frac{p}{\sqrt{3}\pi}g\cdot \bar{f}.
\eeq 
This is the well-known STM integral equation for bosons ($p=+1$) and fermions ($p=-1$) for the half-off-shell, amputated, connected Greens function\footnote{up to our redefinition (\ref{bf1})} \cite{Braaten04}.

\begin{figure}[t]
\includegraphics[width=120mm]{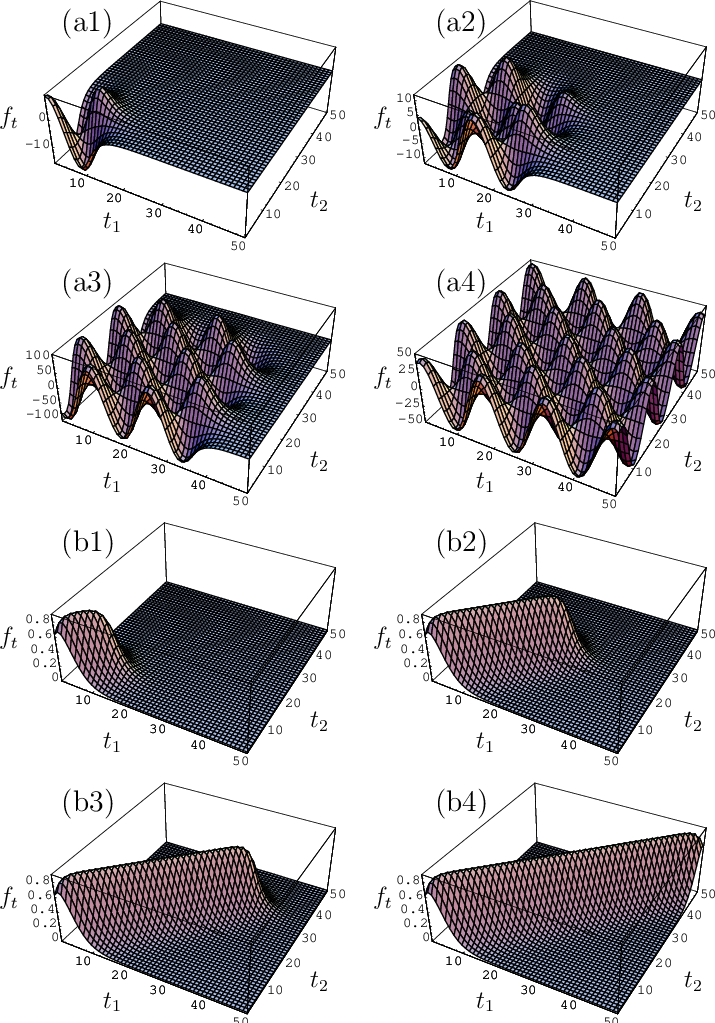}
\caption{\small{The RG evolution of the momentum dependent modified vertex $f_{t}(t_1,t_2)=4(3+p)q_{1}q_{2}\lambda_{3}(q_{1},q_{2},E)$ for bosons (a1-a4) and SU(2) fermions (b1-b4). Spatial momenta $t_1$, $t_2$ and the RG time $t$ are descritized to $N=50$ intervals with a step $\Delta t= 0.4$. The cartoons for bosons and fermions correspond to the descritized steps $10, 25, 35, 50$.
}}
\label{cartoon}
\end{figure}

\begin{figure}[t]
\includegraphics[width=140mm]{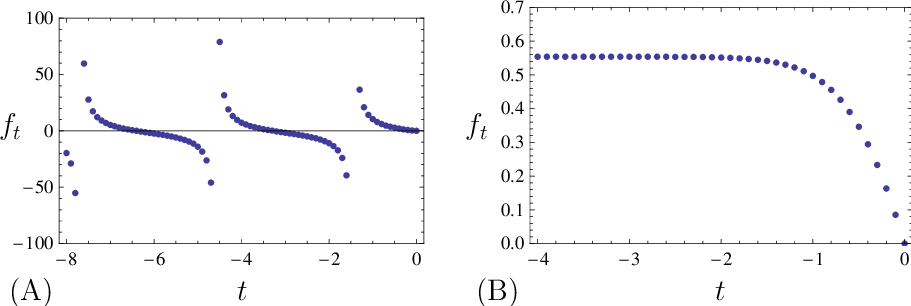}
\caption{\small{Numerical evolution in the RG time $t$ of $f_{t}(t_{1}=0,t_{2}=0)$ for System I (A) and System II (B). For SU(2) fermions (B) the modified vertex approaches a fixed point solution; in the case of bosons (A), a limit cycle behavior is developed with a period $T_{temp}\approx 3.1$. 
}}
\label{RGtime}
\end{figure}

The difference of the signs in Eq. (\ref{bf7}) between System I and II turns out to be crucial. In order to see that, we solve Eq. (\ref{bf7}) numerically by discretization. A series of cartoons of the evolution of the reduced 1PI vertex $f_{t}(t_{1},t_{2})$ for both systems is shown in FIG. \ref{cartoon}. For fermions, first a peak appears in the UV ($t_{1}=0,t_{2}=0$), which propagates in the diagonal direction ($t_{1}=t_{2}$) during the RG evolution. On the other hand, for bosons, a periodic structure (with period $T_{spatial}\approx 6.2$ in both directions) develops gradually. Now it is clear why the approximation investigated in the last section failed to give the quantitatively correct result. The pointlike approximation, which corresponds to a planar landscape (no $t_{1}$ and $t_{2}$ dependence, see Section \ref{triatom}), is not valid in the three-body sector (for more details see Appendix \ref{separable}). The evolution in the RG time $t$ of the UV point $f_{t}(t_{1}=0,t_{2}=0)$ for both systems is depicted in FIG. \ref{RGtime}. While for fermions the evolution is monotonic in time, in the case of bosons we obtain a ``temporal'' oscillation of period $T_{temp}\approx 3.1$. For different points in the $t_{1}-t_{2}$ plane the ``time'' evolution is triggered at the scale $t_{in}\sim O(t_{1},t_{2})$.

The numerical solution for bosons is consistent with the results of \cite{Bedaque1, Bedaque2}. Spatial and temporal oscillations are correlated. As found in \cite{Bedaque1, Bedaque2} evolution in the RG time develops a limit cycle behavior. The Efimov parameter $s_{0}$ can be calculated $s_{0}=\frac{\pi}{T_{temp}}\approx 1.0$ (see Section \ref{triatom}) which is in a good agreement with the Efimov result $s_{0}\approx1.00624$. The accuracy of our result is limited by the numerical procedure only.


\section{Three-body sector: System III}
\label{thg}
As introduced in Sect. \ref{Threeflow} for $SU(3)$ fermions there are two specific situations (System III(a) and System III(b)), when there is a single flow equation instead of the general two. Fortunately, both cases can formally be solved for $E=0$ in a similar fashion compared to Sect. \ref{one-two}. In fact, System III(a) is completely equivalent to System II (see TABLE I) such that we obtain a fixed point solution in this case (see FIG. \ref{RGtime}). For System III(a) we follow similar steps as in Sect. \ref{one-two}: we define a reduced atom-diatom 1PI vertex $f_{t}(t_1,t_2,E)$ (\ref{bf1}) and obtain a flow equation for the reduced vertex (\ref{bf2}) with the coefficients $A=4$, $B=4$ and $C=1$.  These coefficients form a complete square and thus it is useful to define the fully connected atom-diatom vertex:
\beq \label{thg1}
\bar{f}_{t}(t_{1},t_{2})=f_{t}(t_{1},t_{2})+2g(t_{1},t_{2}).
\eeq
The flow equation (\ref{bf2}) now reads:
\beq \label{thg2}
\partial_{t}\bar{f}_{t}(t_{1},t_{2})=-\frac{1}{\sqrt{3}\pi}\bar{f}_{t}(t_{1},t)\bar{f}_{t}(t,t_{2}), \qquad \bar{f}_{t=0}(t_{1},t_{2})=2g(t_{1},t_{2}).
\eeq
The equation and the initial condition are identical to Eq. (\ref{bf4}) for bosons\footnote{The initial condition for $SU(3)$ fermions is $\bar{f}_{t=0}(t_{1},t_{2})=2g(t_{1},t_{2})$, while for bosons one has $\bar{f}_{t=0}(t_{1},t_{2})=g(t_{1},t_{2})$. However, for bosons $g(t_1,t_2)$ is two times larger than for fermions (\ref{bf1}). Thus the initial conditions are identical.}. For this reason we expect the appearance of the Efimov effect for the $SU(3)$ fermionic System III(b) with  the Efimov parameter $s_0\sim1.00624$.

At first sight it seems surprising that both bosons and $SU(3)$ fermions have the identical Efimov parameter $s_0$. As explanation, we propose a simple possible quantum mechanical argument: In order to find a bound state spectrum for SU(3) fermions one must solve the three-body Schr\"odinger equation. The total wave function must be totally \emph{antisymmetric} for fermions. We can achieve this by taking the total wave function as the product of a totally antisymmetric flavor part ($\epsilon_{ijk}|i>|j>|k>$) times a totally \emph{symmetric} orbital part. Hence the orbital part has the same symmetry property as the bosonic case. Only the orbital part is needed for the quantum mechanical calculation of the bound state problem, which leads to the identical Efimov parameters for bosons and SU(3) fermions. 
\section{Conclusions and Outlook}
\label{Conclusion}
This work applies the method of functional renormalization to the few-body physics of atoms near a Feshbach resonance. We investigate three different systems, namely identical bosons as well as two and three species of fermions. The two-body sector is solved exactly. The unitarity limit is governed by a fixed point and all three systems seem to be examples of the non-relativistic conformal field theories. In the three-body sector, however, no infrared fixed point exists for bosons and three-component fermions. We solve the momentum-dependent problem of the three-body sector at unitarity. This leads to the Skorniakov-Ter-Martirosian equation, well-known from quantum mechanics. A numerical solution for $U(1)$ bosons and $SU(3)$ fermions shows the emergence of the Efimov effect; the appearance of an infinite geometric spectrum of triatom states. Hence in these systems the continuous scaling symmetry is broken to the discrete scaling subgroup $Z$ by a quantum anomaly. The renormalization group flow develops a limit cycle behavior (see FIG. \ref{RGtime}). The Efimov parameter $s_{0}$ for the three-component fermions is found to be identical to the Efimov parameter of the well-studied bosonic case, which agrees with the quantum-mechanical prediction.

The current work can be extended in various ways: One can go away from unitarity in the three-body sector and derive universal properties such as recombination rates and the positions of diatom-triatom thresholds. Our technique allows us to investigate equilibrium states with non-zero density and temperature. This can be achieved by simply changing the chemical potential and introducing the temperature by replacing the $\omega$-integrals by the discrete sums of the Matsubara formalism. In that case the effective, approximate description of the models in terms of various composite fields (e.g. trions, density bosons) would be very useful due to a large reduction of the numerical effort. The description of a simple, but efficient effective theory is summarized in Appendix \ref{bosonization}. The excellent agreement of the vacuum solution with high precision quantum-mechanical computations provides a robust starting point for the investigation of the many-body system of nonzero density and temperature.

\emph{Acknowledgments} -- S.M is especially grateful to S. Diehl for enlightening discussions and important suggestions. We acknowledge the discussions with J. M. Pawlowski, H. Gies, H. C. Krahl, M. Scherer, S. Jochim, T. B. Ottenstein, T. Lompe, M. Kohnen and A. N. Wenz. We are indebted to J. Hosek and H.-W. Hammer for providing important remarks.

\appendix
\section{Galilean and non-relativistic conformal symmetry} \label{NRCFT}
All systems we consider in this work (\ref{M3}), (\ref{M4}) and (\ref{M5}) have a centrally extended Galilean spacetime symmetry\footnote{This is the non-relativistic analogue of the Poincare group in relativistic QFT.}. The centrally extended Galilean algebra consists of eleven generators: particle number $N$ (central charge), time translation $H$, three spatial translations $P_{i}$, three spatial rotations $M_{ij}$ and three Galilean boosts $K_{i}$. The non-trivial commutators are (in the real time formalism):
\beq \label{NRCFT1}
[M_{ij},M_{kl}]=i(\delta_{ik} M_{jl}-\delta_{jk} M_{il}+\delta_{il} M_{kj}-\delta_{jl} M_{ki}) ,
\eeq
\beq \label{NRCFT2}
[M_{ij},K_{k}]=i(\delta_{ik} K_{j}-\delta_{jk} K_{i}), \qquad [M_{ij},P_{k}]=i(\delta_{ik}P_{j}-\delta_{jk} P_{i}),
\eeq
\beq \label{NRCFT3}
[P_i,K_j]=-i\delta_{ij}N, \qquad [H,K_{j}]=-iP_{j}.
\eeq
In the case of a free non-relativistic field theory the group of spacetime symmetries is in fact larger than the Galilean group \cite{Hagen, Niederer} and is called the Schr\"odinger group\footnote{This is the non-relativistic counterpart of the conformal group.}. For the dynamical exponent $z=2$ there are two additional generators: the scaling generator $D$ and the special conformal generator $C$. The scale symmetry acts on the time and spatial coordinates according to:
\beq \label{NRCFT3a} (x_{i},t) \to (x^{\prime}_{i},t^{\prime})=(\lambda x_i,\lambda^2 t),  \eeq 
where $\lambda$ is a scale parameter. A special conformal transformation on time and spatial coordinates is given by \cite{Hagen}:
\beq \label{NRCFT3b} (x_{i},t) \to (x^{\prime}_{i},t^{\prime})=(\frac{x_i}{1-c t}, \frac{t}{1-c t}), \eeq
where $c$ is a parameter of the special conformal transformation.
The additional, non-trivial commutators of the Schr\"odinger algebra are:
\beq \label{NRCFT4}
[P_i,D]=-iP_i, \quad [P_i,C]=-iK_i, \quad [K_i,D]=iK_i,
\eeq 
\beq \label{NRCFT5}
[D,C]=-2iC, \quad [D,H]=2iH, \quad [C,H]=iD.
\eeq 
It is important to note that besides the free theory there are few known examples of interacting theories which are symmetric with respect to the Schr\"odinger group. These theories are called non-relativistic conformal field theories (NRCFT) and $SU(2)$ non-relativistic fermions at unitarity (System II) are believed to constitute one of them. 

In analogy to relativistic conformal field theories it is possible to introduce primary operators in an NRCFT \cite{Nishida}. A local primary operator $\mathscr{O}(t,\mathbf{x})$ has a well defined scaling dimension $\Delta_{\mathscr{O}}$ and particle number $N_{\mathscr{O}}$:
\beq \label{NRCFT6}
[D,\mathscr{O}]=i\Delta_{\mathscr{O}} \mathscr{O}, \qquad [N,\mathscr{O}]=N_{\mathscr{O}}\mathscr{O},
\eeq
where $\mathscr{O}\equiv\mathscr{O}(t=0,\mathbf{x}=0)$. The primary operator $\mathscr{O}$ also commutes with $K_i$ and $C$:
\beq \label{NRCFT7}
[K_{i},\mathscr{O}]=0 \qquad [C,\mathscr{O}]=0.
\eeq
It is possible to show that the operators, constructed by taking spatial and time derivatives of a primary operator $\mathscr{O}$, form an irreducible representation of the Schr\"odinger group. Similar to the relativistic case the form of the two-body Greens function of the primary operators is fixed by the conformal symmetry (in the imaginary time formalism) \cite{Nishida}:
\beq \label{NRCFT8}
<\mathscr{O}\mathscr{O}^{\dagger}>\sim (i\hat{\omega}+\frac{\mathbf{q}^2}{2M N_{\mathscr{O}}})^{\nu},
\eeq
where $\nu=\Delta_{\mathscr{O}}-5/2$ for $d=3$. The simplest examples of primary operators in the theory of SU(2) symmetric fermions are the atom operator $\psi$ ($N_{\psi}=1, \Delta_{\psi}=3/2$) and the diatom operator $\varphi$ ($N_{\varphi}=2, \Delta_{\varphi}=2$).  The form of the inverse propagators at unitarity, which we found to be given by Eq. (\ref{t13}), is consistent with Eq. (\ref{NRCFT8}).
\section{Completion of the vertex expansion to $\Gamma^{(4)}_{k}$} \label{Gamma4}
In this appendix we complete the vertex expansion to fourth order and argue that our truncations (\ref{M3},\ref{M4},\ref{M5}) are sufficient to perform exact calculations for the three-body sector. At fourth order in the fields there are only two more vertices, which are compatible with the internal symmetries of the considered models:
\begin{eqnarray} \label{gamma1}
&\Gamma(4)_{\psi}=\frac{1}{2}\int\limits_{Q_1,...Q_4}  \lambda_{\psi}(Q_{1},Q_{2},Q_{3})\psi^{\dagger}(Q_{1})\psi(Q_{2})\psi^{\dagger}(Q_{3})\psi(Q_{4})\delta(-Q_{1}+Q_{2}-Q_{3}+Q_{4}), & \nonumber \\
&\Gamma(4)_{\varphi}=\frac{1}{2}\int\limits_{Q_1,...Q_4}  \lambda_{\varphi}(Q_{1},Q_{2},Q_{3})\varphi^{\dagger}(Q_{1})\varphi(Q_{2})\varphi^{\dagger}(Q_{3})\varphi(Q_{4})\delta(-Q_{1}+Q_{2}-Q_{3}+Q_{4}).& 
\end{eqnarray}
In the two-channel model considered in this work, we choose the initial UV value of the vertex $\lambda_{\psi}$ to be zero, $\lambda_{\psi}=0$.  This means that the interaction between the atoms are described at the microscopic level by the exchange of diatom states. With $\lambda_\psi=0$ at the UV scale this coupling is not regenerated by the flow in vacuum. The  one-loop diagrams contributing to the flow have inner lines pointing in the same direction with respect to the loop momentum and therefore vanish in vacuum (see Sect. \ref{VacPro}). Thus $\lambda_{\psi}=0$ is a fixed point. The flow away from this fixed point has been studied for the system with two species  of fermions in \cite{DGPW2}.

The 1PI vertex $\lambda_{\varphi}$ belongs to the four-body sector (for definition of the n-body sector see Sect. \ref{VacPro}) and it decouples from the flow equations of the two and three-body sectors due to the vacuum hierarchy (for more details see Sect. \ref{VacPro}). Thus our truncations (\ref{M3},\ref{M4},\ref{M5}) are sufficient to obtain the exact vacuum physics of the three-body sector.
\section{Bound state approximation and separable potential}
\label{separable}

In this appendix we present an alternative solution of the two-body sector using the Lippmann-Schwinger equation of quantum mechanics, which helps to elucidate the efficiency of the two-channel model and the limitations 
of the trion approximation in \cite{FSMW}.

The one-channel model provides an alternative description of ultracold atoms near a broad Feshbach resonance. This model contains the atom field $\psi$ only and the microscopic action is given by\footnote{For simplicity, we present the one-channel model for $U(1)$ bosons at \emph{unitarity} only. However, our arguments can be extended away from unitarity and are applicable to both $SU(2)$ and $SU(3)$ fermion systems.}:
\beq \label{bs1}
\Gamma_{t=0}=\int\limits_{Q}\psi^{*}(Q)(i\omega +q^2)\psi(Q)+\frac{\lambda_{\psi}}{2}\int\limits_{Q_1,...,Q_4}\psi^{*}(Q_1)\psi(Q_2)\psi^{*}(Q_3)\psi(Q_4)\delta(-Q_1+Q_2-Q_3+Q_4),
\eeq
where $\lambda_{\psi}$ is a pointlike four-atom interaction which is related to the s-wave scattering length in the IR. Roughly speaking, the quantum-mechanical atom-atom interaction potential of the one-channel model (\ref{bs1}) in Minkowski space is given by\footnote{Strictly speaking, the contact interaction is ill-defined and must be regularized. This can be done by introducing the pseudo-potential $V(\mathbf{r})\psi(\mathbf{r})=\lambda_{\psi}\delta^{(3)}(\mathbf{r})\frac{\partial}{\partial r}(r\psi(\mathbf{r}))$. Here we use an alternative regularization by introducing a momentum cutoff $\Lambda$ directly into the Lippmann-Schwinger equation.}:
\beq \label{bs2}
V(\mathbf{x})=\frac{\lambda_{\psi}}{2}\delta^{(3)}(\mathbf{x}).
\eeq  
Let us now perform a Fourier transformation of this potential:
\beq \label{bs3}
V(\mathbf{k},\mathbf{k}^{\prime})=\int d^{3} r \exp[-i (\mathbf{k}^{\prime}-\mathbf{k})\cdot \mathbf{r}] V(\mathbf{r})=\frac{\lambda_{\psi}}{2}.
\eeq
At this point two important remarks about the potential (\ref{bs3}) can be made:
\begin{itemize}

\item $V(\mathbf{k},\mathbf{k}^{\prime})$ is a separable potential because it can be written in the form $\frac{\lambda_{\psi}}{2}U(\mathbf{k})U(\mathbf{k}^{\prime})$.
\item $V(\mathbf{k},\mathbf{k}^{\prime})$ is $\mathbf{k}$ and $\mathbf{k}^{\prime}$ independent, i.e. $U(\mathbf{k})=U(\mathbf{k}^{\prime})=1$.

\end{itemize}
We investigate the atom-atom scattering in the center-of-mass frame. The Lippmann-Schwinger integral equation for the K-matrix is \cite{Wilson}:
\beq \label{bs4}
K(\mathbf{k},\mathbf{k}^{\prime}, E)=V(\mathbf{k},\mathbf{k}^{\prime})+\mathscr{P}\int\limits^{\Lambda}\frac{d^3 q}{(2\pi)^3}\frac{V(\mathbf{k},\mathbf{q})K(\mathbf{q},\mathbf{k}^{\prime},E)}{E-2\mathbf{q}^2},
\eeq
where $\mathscr{P}$ denotes the Cauchy principle value and $\Lambda$ is a momentum cutoff, which regularizes the contact interaction. The K-matrix is similar to the T-matrix but uses a standing wave boundary condition which leads to the principal value prescription in Eq. (\ref{bs4}). The kinematics of $K(\mathbf{k},\mathbf{k}^{\prime},E)$ is similar to the kinematics depicted in FIG. \ref{fig:2}.
The integral equation (\ref{bs4}) can be easily solved in the special case of a \emph{separable potential}. The solution factorizes:
\beq \label{bs5}
K(\mathbf{k},\mathbf{k}^{\prime}, E)=-\frac{U(\mathbf{k})U(\mathbf{k}^{\prime})}{D(E)},
\eeq
where $D(E)$ is given by:
\beq \label{bs6}
D(E)=-\frac{2}{\lambda_{\psi}}+\mathscr{P}\int\limits^{\Lambda}\frac{d^{3}q}{(2\pi)^3}\frac{U^{2}(\mathbf{q})}{E-2\mathbf{q}^2}.
\eeq
In the special case of the contact interaction Eq. (\ref{bs5}) depends only on E. This means that the exact atom-atom scattering amplitude in the center-of-mass frame can be rewritten in terms of the exchange of the composite diatom with inverse propagator $P_{\phi}(E,\mathbf{p}=0)\sim D(E)$. For the contact interaction $D(E)$ is given by:
\beq \label{bs7}
D(E)=-\frac{2}{\lambda_{\psi}}+\mathscr{P}\int^{\lambda}\frac{dq}{2\pi^2}\frac{q^2}{E-2q^2}=-\frac{2}{\lambda_{\psi}}-\frac{\Lambda}{4\pi^2}+\frac{E}{8\pi^2}\mathscr{P}\int^{\Lambda}\frac{dq}{q^2-E/2}.
\eeq
The microscopic $\lambda_{\psi}$ can be adjusted such that
\beq
\frac{2}{\lambda_{\psi}}+\frac{\Lambda}{4\pi^2}\sim a^{-1}.
\eeq
At unitarity the first two terms in the second Eq. (\ref{bs7}) cancel. The last integral is convergent, hence we take $\Lambda\to\infty$. By dimensional analysis we obtain:
\beq \label{bs8}
D(E)\sim \sqrt{E}.
\eeq
To summarize, the atom-atom scattering amplitude is momentum-independent in the center-of-mass frame. Hence the two-body sector can be solved exactly by introducing a diatom exchange in the s-channel. By Galilean symmetry this result can be extended to a general reference frame:
\beq \label{bs9}
D(E,\mathbf{k})\sim \sqrt{E-\frac{\mathbf{k}^2}{4 M_{\psi}}}
\eeq
The functional form of the inverse diatom propagator is consistent with our findings (\ref{t13}) in Sect. \ref{twobody}.

In the three-body channel the atom-diatom interacting potential is momentum dependent in the center-of-mass frame. The momentum dependence is generated by the box diagram (see first diagram on RHS of FIG. \ref{fig:3}). For this reason the trion approximation, which we used in \cite{FSMW}, and in particular the pointlike approximation (Sect. \ref{triatom}), do not fully capture this momentum dependence and is not as efficient as the ``diatom trick''. It leads to the quantitative inaccuracy of the Efimov parameter $s_{0}$.
\section{Analysis of $\frac{d}{dt}f(t)=\alpha f(t)^2+\beta f(t)+\gamma$}
\label{triatomap}

\begin{figure}[t]

\includegraphics[width=120mm]{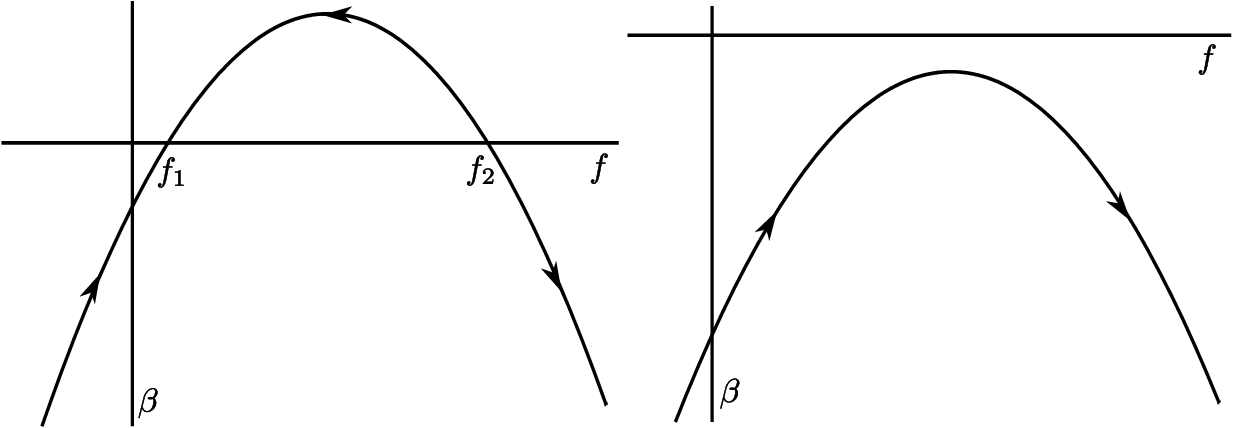}

\caption{\small{$\beta$-function of the RG equation (\ref{ad1}) for $D>0$ (A) and $D<0$ (B). The arrows show the direction of the RG flow toward the IR.
}}
\label{fig:ap}
\end{figure}

In this appendix we perform an analysis of the differential equation, which we encountered in the calculation of the three-body sector in the pointlike approximation:
\beq \label{ad1}
\frac{d}{dt}f(t)=\alpha f(t)^2+\beta f(t)+\gamma, \qquad f(t_{0})=f,
\eeq
with $\alpha,\beta,\gamma\in\mathbb{R}$. The form of the solution is determined by the sign of the discriminant of the $\beta$-function $D\equiv \beta^{2}-4\alpha \gamma$. There are three different cases (without loss of generality we take $\alpha\le0$, which is the case for Systems I, II and III):
\begin{itemize}
\item $D > 0$

In this case the $\beta$-function has two fixed points $f_{1}$ (IR stable) and $f_{2}$ (IR unstable) with $f_1<f_2$ (see FIG. \ref{fig:ap}). For the initial condition $f<f_{1}$ the solution is attracted to the fixed point $f_{1}$ in the IR. If the initial condition is $f>f_{2}$, the solution is repelled from the fixed point $f_{2}$ and in driven towards a Landau pole. Formally, we can extend the solution beyond the Landau pole. For an initial condition $f_{1}<f<f_{2}$ the solution is attracted by the IR fixed point $f_{1}$ and has the form:
\beq
f(t)=\frac{-\beta+\sqrt{D}\tanh \left[\frac{\sqrt{D}}{2}(t+\kappa)\right]}{2\alpha},
\eeq  
with $\kappa$ fixed by the initial condition.
\item $D=0$

In this case there is one fixed point and we have four distinct subcases:
\renewcommand{\labelenumi}{\alph{enumi})}
\begin{enumerate}
\item $\alpha=\beta=\gamma=0$ $\Rightarrow$ A trivial solution $f(t)=f$ is obtained.
\item $\alpha=\beta=0$ and $\gamma\ne0$ $\Rightarrow$ We have a linear solution $f(t)=f+\gamma (t-t_{0})$.
\item $\alpha\ne 0$ and $ \beta= \gamma=0$ $\Rightarrow$ We obtain the Landau pole solution:
\beq \label{}
f(t)=\frac{f}{1+f\alpha(t-t_{0})}.
\eeq
This solution can be formally extended beyond the Landau pole.
\item $\alpha\ne 0$, $ \beta\ne 0$ and $ \gamma \ne 0$ $\Rightarrow$ This case can be put in the form of the case c) by a transformation $f\to f+\sqrt{\frac{\gamma}{\alpha}}$.
\end{enumerate}

\item $D < 0$

In this case there are no fixed points. The formal solution can be written as:
\beq
f(t)=\frac{-\beta+\sqrt{-D}\tan \left[\frac{\sqrt{-D}}{2}(t+\kappa)\right]}{2\alpha},
\eeq
where $\kappa$ is fixed by the initial condition. This solution is periodic with a period $T=\frac{2\pi}{\sqrt{-D}}$.
\end{itemize}

It is important to mention that the discriminant $D$ is invariant under the multiplicative reparametrization $f(t)\to \omega f(t)$ with some constant $\omega$. This means that both the classification and the period $T$ are not sensitive to the multiplicative reparametrization of $\lambda_{3}^{R}$ in Sect. \ref{triatom}.

\section{Bosonization and Fermionization} \label{bosonization}
It was shown in Sects. \ref{one-two} and \ref{thg} that for a quantitative precise description of the three-body physics the momentum dependence of the vertex $\lambda_3$ is quite important. Although it was eventually possible to take this momentum dependence into account, we had to pay a price for it. Solving the flow equation numerically while taking the full s-wave projected momentum dependence of the vertices into account needs a relatively large numerical effort. While this is still manageable in the vacuum where both the density and the temperature vanish, the numerical cost would be significantly larger in the more general case of nonzero density or temperature. It is therefore reasonable to look for an effective approximate description that is numerically less expensive but nevertheless leads to good numerical precision. 

For this task it is crucial to find a simple way to take at least the qualitative features of the momentum dependence into account. For simplicity we concentrate this discussion on System III only. Let us start with the discussion of $\lambda_{3b}$. From Eq. \eqref{M5} we can read off that it corresponds to a channel where the fermionic and bosonic spins are contracted separately
\begin{equation}
-\lambda_{3b}\,\psi^*_i\psi_i\, \varphi^*_j \varphi_j.
\end{equation}
One might describe this vertex by the exchange of a real two-component boson field $\sigma=(\sigma_{\psi},\sigma_{\varphi})$ which couples to the composite operators $\psi^\dagger \psi$ and $\varphi^\dagger \varphi$ with some Yukawa-type interactions. More explicit, we could use the following action:
\begin{eqnarray}
\nonumber
\Gamma_\sigma=\int_Q \sigma_{\psi}(Q) P_\sigma(Q)\sigma_{\varphi}(-Q)-\int_{Q_1,Q_2}h_{\sigma \psi}(Q_1,Q_2)\psi^\dagger(-Q_1)\psi(Q_2)\sigma_{\psi}(Q_1+Q_2)\\
-\int_{Q_1,Q_2}h_{\sigma \varphi}(Q_1,Q_2)\varphi^\dagger(-Q_1)\varphi(Q_2)\sigma_{\varphi}(Q_1+Q_2).
\label{sigmaaction}
\end{eqnarray}
Since $\sigma$ is a real boson, its propagator fulfills $P_\sigma(Q)=P_\sigma(-Q)$. Together with Galilean invariance this implies that $P_\sigma$ does not have any frequency dependence $P_\sigma(Q)=P_\sigma(\mathbf{q})$ \cite{Henkel}. The exchange of a $\sigma$-boson corresponds to an instantaneous interaction. Note, however, that Galilean invariance is broken spontaneously by a condensate or a Fermi surface at nonzero density. In that case the $\sigma$-boson becomes dynamical and corresponds to a propagating phonon. An expectation value of $\sigma$ corresponds to a shift in the effective chemical potential \cite{Diehl:2005ae}.

In Eq. \eqref{sigmaaction} the field $\sigma$ appears quadratic and we can eliminate it by solving its field equation. This results in:
\begin{eqnarray}
\nonumber
\Gamma_\sigma =
-\int_{Q_1..Q_4}\frac{h_{\sigma\psi}(Q_1,Q_2)h_{\sigma\varphi}(Q_3,Q_4)}{P_\sigma(Q_1+Q_2)}\,\psi^\dagger(-Q_1)\psi(Q_2)\varphi^\dagger(-Q_3)\varphi(Q_4) \,\delta(Q_1+Q_2+Q_3+Q_4).
\end{eqnarray}
We observe that we get a tree-level contributions that has the spin-structure of the term $\sim \lambda_{3b}$. Assuming Yukawa couplings that are independent of the momenta and using the conventions of Sect. \ref{Threeflow} the contribution to $\lambda_{3b}$ reads
\begin{equation}
\lambda_{3b,\sigma-\text{exchange}}(\mathbf{p_1},\mathbf{p_2};E)=\frac{ h_{\sigma\psi}h_{\sigma\varphi}}{P_\sigma(\mathbf{p_1}-\mathbf{p_2})}.
\end{equation}
For an inverse propagator of the form $P_\sigma(q)=m_\sigma^2+\mathbf{q}^2$ we find after the s-wave projection:
\begin{equation}
\lambda_{3b,\sigma-\text{exchange}}(p_1,p_2;E)=\frac{h_{\sigma\psi}h_{\sigma\varphi}}{4p_1p_2}\text{ln}\left(\frac{p_1^2+p_2^2+2p_1p_2+m_\sigma^2}{p_1^2+p_2^2-2p_1p_2+m_\sigma^2}\right).
\end{equation}
The parameters $m_\sigma^2$ and $h_{\sigma\psi}, h_{\sigma\varphi}$ can be chosen such that the form of $\lambda_{3b}$ is resembled closely. One can compare this to the tree-level contribution to $\lambda_{3b}$ by the exchange of a fermion $\psi$. It is obtained from Eq. \eqref{M5} by solving the field equation for $\psi$ and while using the fact that the propagator for $\psi$ and the Yukawa coupling $h$ are not renormalized:
\begin{equation}
\lambda_{3b,\psi-\text{exchange}}(\mathbf{p_1},\mathbf{p_2};E)=\frac{2h^2}{\mathbf{p_1}^2+\mathbf{p_2}^2+(\mathbf{p_1}+\mathbf{p_2})^2-E-\mu_{\psi}}.
\end{equation}
After s-wave projection, this reads:
\begin{equation}
\lambda_{3b,\psi-\text{exchange}}(p_1,p_2;E)=\frac{h^2}{2p_1p_2}\text{ln}\left(\frac{p_1^2+p_2^2+p_1p_2-(\mu_{\psi}+E)/2}{p_1^2+p_2^2-p_1p_2-(\mu_{\psi}+E)/2}\right).
\end{equation}
One can see, that the functional form of the two tree-level contributions after s-wave projection is quite similar. This is also the form of the momentum dependence found in the numerical solution of the flow equation for $\lambda_{3b}$ without the $\sigma$-boson (Sect. \ref{one-two} and \ref{thg}). We therefore expect that the description of $\lambda_{3b}$ as the exchange of a $\sigma$-boson with momentum-independent Yukawa couplings $h_{\sigma\psi}$ and $h_{\sigma_\varphi}$ leads to results that are comparable to the inclusion of the full (s-wave projected) momentum dependence. However, this description would be much more efficient with respect to the numerical effort. For the translation between the description used in the main part of this paper, where $\lambda_{3b}$ is included as an own vertex and the description of $\lambda_{3b}$ in terms of the exchange of a $\sigma$-boson, one might use the method of rebosonization \cite{Gies}.

The vertex $\lambda_{3a}$ in Eq. \eqref{M5} can also be described by the exchange of some particle, which corresponds in this case to a bound state of three atoms, the trimer or trion \cite{FSMW}. Although the vertex $\lambda_{3b}$ and the momentum dependence of the Yukawa-like couplings were neglected in \cite{FSMW}, the behavior found there was already qualitatively correct. Why this triatom approximation is not sufficient to describe the complete momentum dependence of $\lambda_{3}$ is discussed in Appendix \ref{separable}. 

We conclude that an effective description of the three-body physics with only a few couplings seems possible and would facilitate the study of systems at nonzero density and temperature.





\end{document}